\documentclass[pdflatex,sn-nature]{sn-jnl}

\usepackage{graphicx}%
\usepackage{multirow}%
\usepackage{amsmath,amssymb,amsfonts}%
\usepackage{amsthm}%
\usepackage{mathrsfs}%
\usepackage[title]{appendix}%
\usepackage{xcolor}%
\usepackage{textcomp}%
\usepackage{manyfoot}%
\usepackage{booktabs}%
\usepackage{algorithm}%
\usepackage{algorithmicx}%
\usepackage{algpseudocode}%
\usepackage{listings}%
\usepackage{verbatim} %
\usepackage{float} %
\usepackage{bm} 

\theoremstyle{thmstyleone}%

\theoremstyle{thmstyletwo}%

\theoremstyle{thmstylethree}%

\begin{document}
\newgeometry{left=30mm,right=30mm,top=30mm,bottom=30mm}

\title[Cluster-Adaptive SQD]{Cluster-Adaptive Sample-Based Quantum Diagonalization for Strongly Correlated Systems}

\author[1]{\fnm{Byeongyong} \sur{Park}}\email{by7816@uos.ac.kr}

\author[1]{\fnm{Sanha} \sur{Kang}}\email{sanha9156@uos.ac.kr}

\author[1]{\fnm{Jongseok} \sur{Seo}}\email{sjs981212@uos.ac.kr}

\author[2]{\fnm{Juhee} \sur{Baek}}\email{bjhee1116@gmail.com}

\author[1,3]{\fnm{Doyeol (David)} \sur{Ahn}}\email{dahn@uos.ac.kr}

\author*[2]{\fnm{Keunhong} \sur{Jeong}}\email{doas1mind@sogang.ac.kr}

\affil[1]{\orgdiv{Center for Quantum Information Processing}, \orgname{University of Seoul}, \orgaddress{\street{163 Seoulsiripdae-ro, Dongdaemun-gu}, \city{Seoul}, \postcode{02504}, \country{Republic of Korea}}}

\affil[2]{\orgdiv{Department of Chemistry}, \orgname{Sogang University}, \orgaddress{\street{35 Baekbeom-ro, Mapo-gu}, \city{Seoul}, \postcode{04107}, \country{Republic of Korea}}}

\affil[3]{\orgname{Singularity Quantum Inc.}, \orgaddress{\street{9506 Villa Isle Drive}, \city{Villa Park, CA}, \postcode{92861}, \country{USA}}}

\abstract{\unboldmath
Sample-based quantum diagonalization (SQD) is a hybrid quantum-classical algorithm for estimating ground-state energies in electronic-structure calculations. 
It uses a quantum processor as a sampler to construct a variational subspace, with Hamiltonian projection and diagonalization performed classically. 
A critical step in SQD is self-consistent particle-number recovery guided by a global reference occupancy vector. 
In strongly correlated systems, however, dominant determinants can be distributed across regions of determinant space, causing this reference to become mixture-averaged and biasing recovery toward mean occupations. 
Here, we introduce cluster-adaptive SQD (CSQD), which clusters pooled single-spin strings and performs particle-number recovery using cluster-specific reference occupancy vectors. 
Under a matched variational budget, CSQD lowers ground-state energies relative to SQD by up to $15.95~\mathrm{mHa}$ for stretched $\mathrm{N_2}$ in a $(10e,26o)$ active space and $57.82~\mathrm{mHa}$ for $\mathrm{[2Fe\!-\!2S]}$ in a $(30e,20o)$ active space. These results suggest that CSQD better captures dispersed occupation structure in strongly correlated systems.
}

\maketitle

\section{Introduction}\label{introduction}
Describing strongly correlated electronic systems, such as transition-metal oxides~\cite{strong_metal}, high-temperature superconductors~\cite{strong_super}, and multiradical species~\cite{strong_radical1, strong_radical2}, remains a central challenge in modern quantum chemistry and condensed matter physics~\cite{strong_hard1, strong_hard2, strong_hard3, strong_hard4, strong_hard5}. 
A characteristic feature of these systems is that the electron--electron Coulomb interaction competes with or even dominates the kinetic energy. 
This competition induces a dense manifold of near-degenerate many-electron states, so that the resulting low-energy eigenstates often exhibit multireference character, with weight broadly dispersed over many determinants~\cite{strong_hard4, strong_hard5, multi}.
Consequently, chemically important information may reside not only in dominant configurations but also in subdominant determinants distributed across the broad support of these states.
Such multireference structure lies beyond the reach of the single-reference approximation, so methods such as Hartree--Fock (HF), coupled-cluster theory, and density functional theory (DFT) often fail qualitatively for these systems~\cite{single_hard1, single_hard2, single_hard3}.

Full configuration interaction (FCI), which treats the entire Hilbert space exactly, is limited by the combinatorial growth of the state space and has so far been demonstrated for systems up to 22 electrons in 22 orbitals $(22e,22o)$~\cite{22e22o} and 26 electrons in 23 orbitals $(26e,23o)$~\cite{26e23o}.
As an efficient alternative, selected configuration interaction (SCI) algorithms~\cite{cipsi, sci, hci, shci}, notably heat-bath configuration interaction (HCI)~\cite{hci,shci}, construct compact variational spaces by iteratively selecting Hamiltonian-connected external determinants whose estimated contributions exceed a prescribed cutoff.
However, this local expansion implicitly assumes that chemically important configurations are reachable through successive additions, requiring intermediate determinants to make contributions large enough to exceed the selection cutoff. 
In strongly correlated systems, where wave-function weight is broadly distributed over many configurations, this assumption can become less valid. 
When important configurations are separated by chains of intermediate determinants with negligible weights, classical selection heuristics may require much tighter cutoffs to recover them and may otherwise yield suboptimal variational spaces~\cite{sci_hard2,sci_hard3,sci_hard4}.

Quantum SCI (QSCI) methods offer the potential to overcome the locality limitations of classical SCI, leveraging superposition and interference to explore the Hilbert space more globally~\cite{qsci1,qsci2,qsci3,sqd,sqd_ex4,extsqd,skqd,sqdrift,afqmc,embedding,pigen,sqd_ex1,sqd_ex2,sqd_ex3}.
In these approaches, the quantum processor serves as a sampler to identify chemically important determinants, while Hamiltonian projection and diagonalization are delegated to classical computation.
Among QSCI workflows, sample-based quantum diagonalization (SQD)~\cite{sqd} has rapidly emerged as a promising framework.
A core feature of SQD is its symmetry-recovery step, which restores particle-number symmetry in noisy samples using a self-consistently updated reference occupancy vector.
Recent demonstrations~\cite{sqd, sqd_ex4} on iron--sulfur (Fe--S) clusters using up to 77 qubits have provided evidence for the feasibility of this strategy in the pre-fault-tolerant quantum computing era.
The SQD protocol is evolving into a broader paradigm for hybrid quantum--classical computing, including extensions such as extended SQD (Ext-SQD)~\cite{extsqd} for excited states and sample-based Krylov quantum diagonalization (SKQD)~\cite{skqd} based on time-evolved Krylov-state sampling.

However, an important limitation of SQD in strongly correlated systems is its reliance on a single global reference occupancy vector for particle-number recovery. 
In such systems, where wave-function weight is broadly dispersed and the occupation structure can be heterogeneous, this reference can become mixture-averaged and fail to represent distinct occupancy patterns.
Particle-number recovery guided by such an averaged reference biases the corrected samples toward a mean occupation pattern, blurring dispersed occupation structure and degrading the determinant pool used for projected diagonalization.

In this work, we introduce cluster-adaptive SQD (CSQD), a reformulation of SQD designed for strongly correlated systems. 
CSQD differs from standard SQD in two key respects. 
First, it treats pooled single-spin strings as the basic units for iterative carry-over and recovery. 
In the carry-over step, pooled single-spin strings are ranked by their aggregate participation across the variational space and propagated to the next iteration, rather than selected according to the weight of any individual full determinant instance. 
This string-level treatment is advantageous when wave-function weight is broadly dispersed, because individual determinant weights may then provide a less reliable measure of the importance of an underlying spin pattern.
Second, CSQD introduces cluster-adaptive particle-number recovery: the pooled single-spin strings are partitioned by unsupervised learning into $K$ clusters, with each cluster assigned its own self-consistently updated reference occupancy vector, and particle-number recovery is performed separately within each cluster using the corresponding reference. 
This cluster-resolved recovery is designed to mitigate the global-reference bias of SQD and its potential failure to capture nonuniform occupation structure.

We benchmarked CSQD against SQD by comparing variational ground-state energy estimates for two representative problems: the dissociation of $\mathrm{N_2}$ in a $(10e,26o)$ active space and the $\mathrm{[2Fe\!-\!2S]}$ cluster~\cite{fes} in a $(30e,20o)$ active space.
For each benchmark instance, we provided both algorithms with identical quantum samples as input and evaluated them under a matched variational budget, so that performance differences reflected only the classical post-processing.
For $\mathrm{N_2}$, SQD often yielded slightly lower variational energies in the weakly correlated regime by at most a few millihartree, whereas in the stretched-bond, strongly correlated regime, CSQD yielded lower variational energies in all but one tested setting, with improvements of up to $15.95~\mathrm{mHa}$ relative to SQD.
For $\mathrm{[2Fe\!-\!2S]}$, the benchmark over 10 self-consistent iterations yielded lower variational energies for CSQD than for SQD across all tested settings, with reductions of up to $45.53~\mathrm{mHa}$ at the largest matched projected-subspace dimension.
An extended 25-iteration follow-up on a representative setting widened the SQD--CSQD gap to $57.82~\mathrm{mHa}$.

To probe the respective roles of the two modifications introduced in CSQD, we additionally examined the $K=1$ single-cluster limit as a control.
Across both benchmarks, this $K=1$ limit generally improved upon SQD in the strongly correlated settings considered, while $K>1$ yielded further gains, indicating that the single-spin-string-based workflow already provides an advantage and that cluster adaptivity contributes an additional improvement.
For $\mathrm{[2Fe\!-\!2S]}$, reference-vector analysis together with post hoc removal-and-refill diagnostics further supported the interpretation that preserving dispersed and nonuniform occupation structure during recovery contributed to the observed energy improvements.

\section{Results}\label{results}
In this section, we first introduce common notation for determinant-based subspace projection and diagonalization, and provide a conceptual overview of CSQD.
Next, we benchmark CSQD against SQD for the dissociation of $\mathrm{N_2}$ as a controlled test case spanning weak to strong correlation.
We then study the $\mathrm{[2Fe\!-\!2S]}$ cluster~\cite{fes} as a challenging strongly correlated benchmark, and use reference-occupancy diagnostics to characterize how the learned reference vectors differ between SQD and CSQD and how this correlates with the observed energy improvements.
To keep this section focused on benchmark outcomes and their interpretation, we report only the methodological and experimental information needed for clarity, while deferring precise algorithmic details, complete hyperparameters, and experimental settings to the Methods section.

\subsection{Problem Setup and Notation}
We consider the electronic-structure problem in a finite basis of spatial orbitals under the Born--Oppenheimer approximation.
Our goal is to estimate the ground-state energy and the corresponding wave function of the electronic Hamiltonian.
In second quantization, the Hamiltonian is given by
\begin{equation}
H
= \sum_{p r,\sigma} h_{p r}\,\hat a^\dagger_{p\sigma}\hat a_{r\sigma}
+ \frac{1}{2}\sum_{p r q s,\sigma\tau} (p r|q s)\,\hat a^\dagger_{p\sigma}\hat a^\dagger_{q\tau}\hat a_{s\tau}\hat a_{r\sigma},
\label{eq:hamiltonian}
\end{equation}
where $p,r,q,s$ label spatial orbitals, $\sigma,\tau\in\{\alpha,\beta\}$ label spin, and $h_{pr}$ and $(pr|qs)$ denote the one- and two-electron integrals.
The operators $\hat a^\dagger_{p\sigma}$ and $\hat a_{p\sigma}$ are the fermionic creation and annihilation operators for the spin orbital $(p,\sigma)$.
The Hamiltonian conserves the numbers of $\alpha$ and $\beta$ electrons; accordingly, we work in a fixed particle-number sector $(N_\alpha,N_\beta)$.
Throughout this work, we restrict to the $S_z=0$ sector with $N_\alpha=N_\beta\equiv N_\sigma$.

We map the fermionic degrees of freedom to qubits using a Jordan--Wigner transformation~\cite{jw}, in which each qubit encodes the occupancy of a spin orbital and computational-basis measurements return bitstrings $x\in\{0,1\}^{2N_{\mathrm{MO}}}$, where $N_{\mathrm{MO}}$ is the number of spatial orbitals.
Each bitstring corresponds to a Slater determinant,
\begin{equation}
|x\rangle=\prod_{p\sigma}\left(\hat a^\dagger_{p\sigma}\right)^{x_{p\sigma}}|\mathrm{vac}\rangle,
\label{eq:determinant}
\end{equation}
where the product is taken in a fixed canonical order of spin orbitals and $|\mathrm{vac}\rangle$ represents the vacuum state.
Subspace diagonalization approximates the ground state by projecting $H$ onto a finite variational subspace $\mathcal{V}$ spanned by a selected set of determinants, and diagonalizing the resulting matrix to obtain a variational energy estimate.
In the following, we use quantum sampling to identify chemically important determinants and construct compact variational subspaces for classical projection and diagonalization.
Accordingly, in each benchmark, a lower projected-Hamiltonian energy at matched final subspace dimension, $\dim(\mathcal{V})$, indicates a higher-quality determinant pool and projected subspace.

\subsection{Overview of CSQD}
\begin{figure}[!t]
\centering
\includegraphics[width=1.0\textwidth]{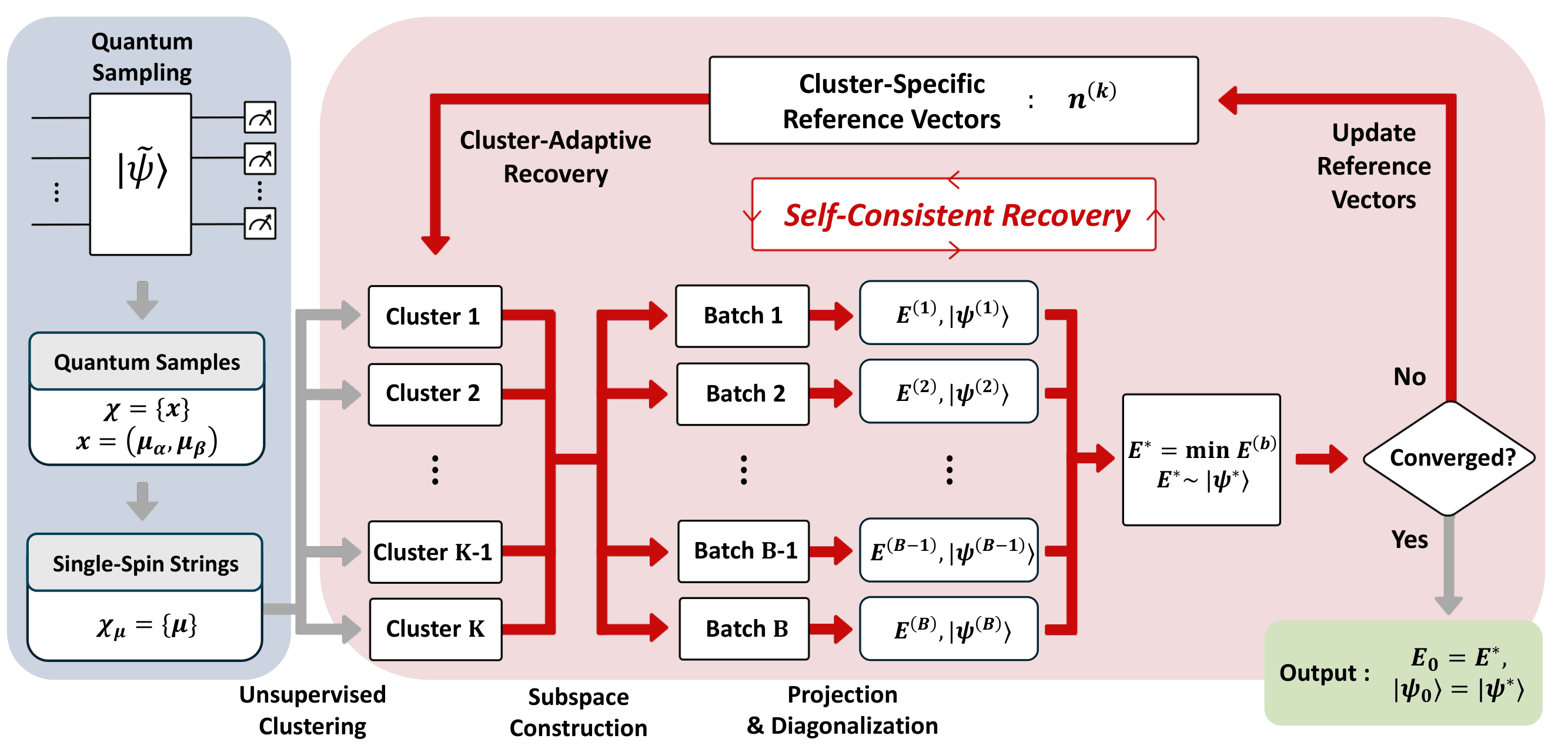}
\caption{\textbf{CSQD workflow.}
The left (blue) block shows quantum sampling and preprocessing. 
Bitstring samples $x=(\mu_\alpha,\mu_\beta)$ obtained from a quantum sampling procedure are pooled as single-spin strings $\mathcal{X}_\mu$.
These strings are partitioned into $K$ clusters by unsupervised clustering.
The right (red) block depicts the self-consistent recovery loop in CSQD.
For each cluster $k$, a cluster-specific reference occupancy vector $\mathbf{n}^{(k)}$ guides particle-number recovery.
The recovered strings are subsampled into $B$ batches, each of which defines a determinant subspace.
For each batch $b$, the Hamiltonian is projected into the corresponding subspace and diagonalized to obtain the lowest-energy eigenpair $\bigl(E^{(b)},|\psi^{(b)}\rangle\bigr)$.
The best batch $b^*=\arg\min_b E^{(b)}$ is then selected, and the reference vectors $\{\mathbf{n}^{(k)}\}_{k=1}^K$ are updated from $|\psi^{(b^*)}\rangle$.
The recovery--diagonalization loop is iterated until a stopping criterion is met, yielding final estimates $(E_0,|\psi_0\rangle)$.}
\label{fig1}
\end{figure}

CSQD is a variant of SQD that treats pooled single-spin strings as the basic units for iterative carry-over and recovery, and performs particle-number symmetry recovery using cluster-specific reference occupancy vectors instead of a single global reference. 
This design is intended to better preserve dispersed and nonuniform occupation structure in strongly correlated systems. 
Fig.~\ref{fig1} provides an overview of the CSQD workflow.

In CSQD, each computational-basis bitstring sample is split into its $\alpha$- and $\beta$-spin components.
The resulting single-spin strings are pooled into a common dataset and partitioned into $K$ clusters via unsupervised learning, such as $k$-modes~\cite{kmodes,kmodes_library} or Bernoulli mixture models (BMM)~\cite{bmm,stepmix}, using empirical frequencies as sample weights.
This clustering enables us to define a cluster-specific reference occupancy vector $\mathbf{n}^{(k)}$ for each cluster $k\in\{1,\dots,K\}$.
Each $\mathbf{n}^{(k)}$ guides particle-number symmetry recovery within cluster $k$.

Next, a shared pool of single-spin strings with the correct particle number is formed by drawing a total of $S$ samples from the clusters in proportion to their statistical weights.
These strings are either intrinsically valid in the initial measurements or corrected via recovery in subsequent iterations.
After deduplication, this pool is truncated to a maximum of $d_{\max}$ strings. 
To mitigate spin contamination arising from a determinant set that is not closed under spin inversion, a spin-symmetric projected subspace is constructed using the shared pool for both the $\alpha$ and $\beta$ sectors, analogous to Ref.~\cite{sqd}.
The determinant basis is then formed as the tensor product of this pool, yielding a projected subspace of dimension up to $d_{\max}^2$. 
The Hamiltonian $H$ is projected into this subspace and diagonalized to obtain a variational ground-state estimate.
To reduce stochastic variability from the subsampling process, subspace construction and projected diagonalization are repeated over $B$ independent batches, and the ground-state solution with the lowest energy is retained.

For each iteration, the cluster-specific reference vectors $\{\mathbf{n}^{(k)}\}$ are updated from the estimated ground state by aggregating squared determinant amplitudes associated with each cluster.
Using the updated reference vectors $\{\mathbf{n}^{(k)}\}$, particle-number symmetry recovery is applied to the initial measurement samples, correcting particle-number-inconsistent single-spin strings separately within each cluster.
To stabilize the subspace across iterations, dominant single-spin strings are carried over to the next iteration.
This entire self-consistent cycle---subspace construction, diagonalization, reference vector update, and recovery---is iterated until the energy and the reference vectors converge, or for at most $T$ iterations.

Relative to SQD, the full CSQD workflow differs in two respects. 
First, subsampling, particle-number recovery, importance evaluation, and carry-over are performed at the level of single-spin strings, with string importance obtained by aggregating squared configuration-interaction (CI) amplitudes over the projected product space. 
Second, in its cluster-adaptive form, CSQD performs recovery using cluster-specific reference occupancy vectors rather than a single global reference. 
As an ablation control, we also consider the single-cluster limit $K=1$ of CSQD, which removes cluster adaptivity while retaining the single-spin-string-based workflow. 
Comparisons with this control help assess the respective contributions of the single-spin-string-based workflow and cluster-adaptive recovery to the observed improvement.
The distinct roles of these two ingredients are discussed in more detail in the Discussion section.

\subsection{Dissociation of $\mathrm{N_2}$: From Weak to Strong Correlation}
The dissociation of $\mathrm{N_2}$ is a standard benchmark that spans the weakly correlated regime near equilibrium and the strongly correlated regime upon bond stretching, thereby providing a comprehensive test of the CSQD workflow.

We studied $\mathrm{N_2}$ in a frozen-core cc-pVDZ basis with an active space of $(10e,26o)$, scanning bond lengths from $R=0.7$ to $3.4~\mathrm{\AA}$ in $0.1~\mathrm{\AA}$ steps.
The equilibrium bond length is $R_e\approx 1.098~\mathrm{\AA}$, and we define the strongly correlated regime as $R\ge 1.5R_e\approx 1.647~\mathrm{\AA}$.
Quantum measurement samples were generated following the SQD protocol of Ref.~\cite{sqd}.
Experiments were performed on the IBM Heron r2 \texttt{ibm\_fez} processor.
For each geometry, we collected $3\times 10^{5}$ shots, leaving $\sim 1.08\times 10^{5}$ valid outcomes on average after reset-mitigation filtering.
Details of the quantum circuit construction, device execution, and error mitigation are provided in the Methods section.

Using the same filtered measurement outcomes as input to both SQD and CSQD, we compared the methods at matched final projected-subspace dimension $\dim(\mathcal{V})=d_{\max}^2$ with $d_{\max}\in\{1000,1500,2000\}$.
For CSQD, we considered both BMM and $k$-modes clustering with $K\in\{2,3,4,5\}$, and additionally performed $K=1$ runs at $d_{\max}=2000$ as a single-cluster control for the CSQD workflow.
CSQD and SQD were run for at most 10 and 11 self-consistent iterations, respectively, using $B=10$ independent batches per iteration.
SQD was granted one additional iteration because its stricter particle-number postselection in the initial subspace construction often yielded a smaller starting pool and hence a smaller initial $\dim(\mathcal{V})$ than CSQD.
In both methods, the string pools used to construct the individual batches were controlled so that the projected-subspace dimension remained close to $d_{\max}^2$ over the self-consistent procedure, ensuring a comparable variational budget.

\begin{figure}[t]
\centering
\includegraphics[width=1.0\textwidth]{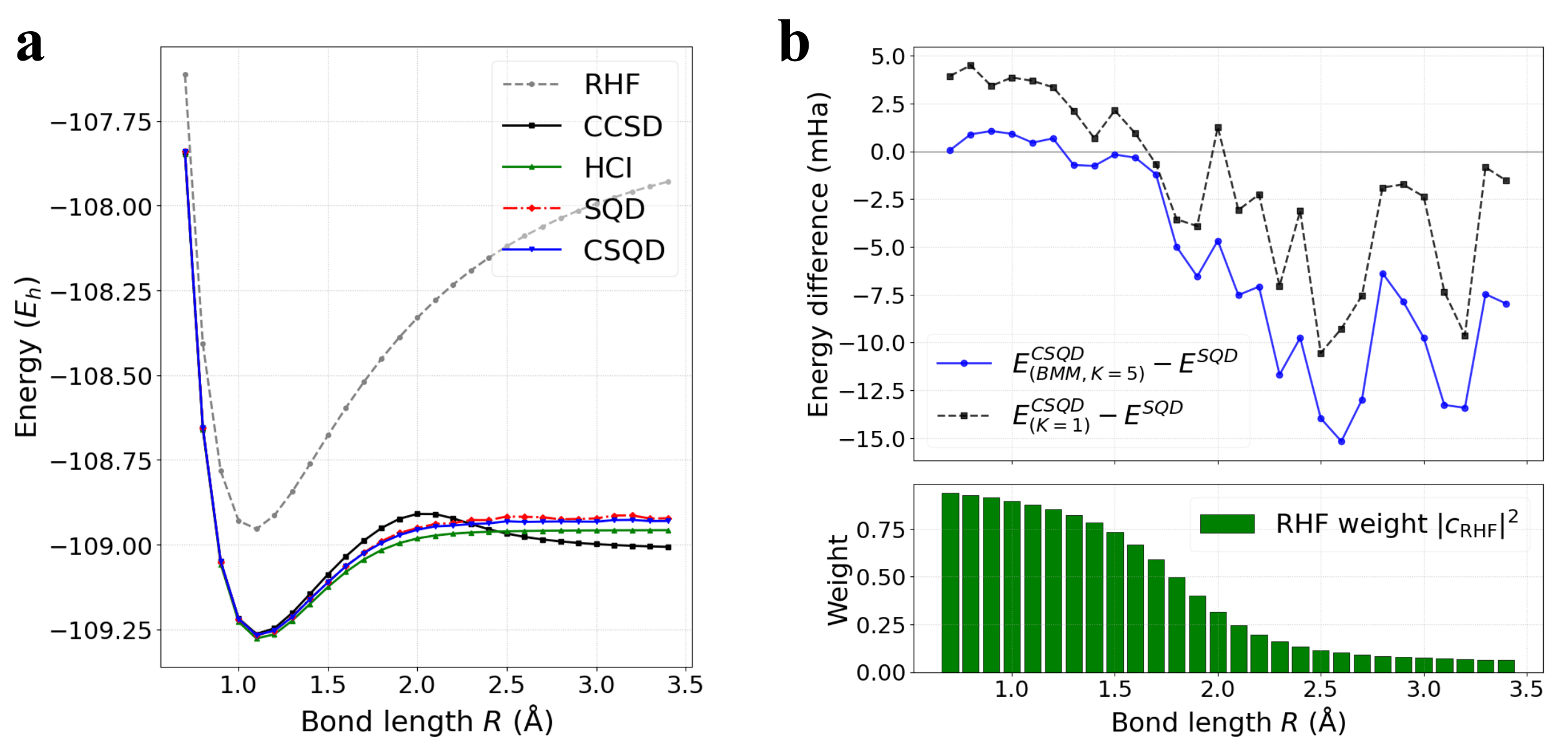}
\caption{\textbf{Dissociation of $\mathbf{N_2}$: Comparison of SQD, CSQD, and classical references.}
In both panels, SQD and CSQD are evaluated from the same quantum inputs and compared at a matched final projected-subspace dimension $\dim(\mathcal V)=d_{\max}^2$ with $d_{\max}=2000$.
\textbf{(a)} Ground-state energy estimates along the $\mathrm{N\!-\!N}$ bond-stretching coordinate $R$ for RHF, CCSD, HCI, SQD, and CSQD (BMM, $K=5$). 
HCI provides a high-accuracy classical benchmark within the chosen active space.
\textbf{(b)} Upper panel: energy differences $E^{\mathrm{CSQD}}_{(\mathrm{BMM},K=5)}(R) - E^{\mathrm{SQD}}(R)$ and $E^{\mathrm{CSQD}}_{(K=1)}(R) - E^{\mathrm{SQD}}(R)$ in mHa. Negative values indicate that CSQD yielded lower variational energies than SQD.
Lower panel: RHF determinant weight $|c_{\mathrm{RHF}}|^2$ (the squared coefficient of the RHF determinant in the HCI expansion), which decreases upon bond stretching, reflecting the growing multireference character. 
} 
\label{fig2}
\end{figure}

In the main text, we focus on the largest value of $d_{\max}$ considered, $d_{\max}=2000$, corresponding to a final projected-subspace dimension of $\dim(\mathcal{V}) = 4\times 10^{6}$.
Fig.~\ref{fig2} visualizes the $\mathrm{N_2}$ dissociation benchmark and highlights the regime-dependent behavior of CSQD for a representative setting (BMM, $K=5$).
BMM is a natural choice because it models binary occupation features through a Bernoulli likelihood.
We also use $K=5$, the largest value considered, as a conservative setting that mitigates potential under-clustering.
The full set of results for smaller $d_{\max}$ values is provided in the Supplementary Information.

In Fig.~\ref{fig2}a, HCI provides a high-accuracy classical benchmark for this active space across both the near-equilibrium and stretched-bond regimes.
Restricted HF (RHF) and coupled-cluster singles and doubles (CCSD) are shown as single-reference baselines.
They perform well near equilibrium but deteriorate upon bond stretching as static correlation becomes increasingly important.
We note that CCSD is not variational and can therefore fall below the benchmark curve in strongly correlated regimes.
Both SQD and CSQD closely track the HCI reference, while CSQD tends to remain closer in the stretched-bond region, where single-reference descriptions become inadequate.

Fig.~\ref{fig2}b makes the SQD--CSQD comparison explicit.
The upper panel shows energy differences $\Delta E_{K=5} = E^{\mathrm{CSQD}}_{(\mathrm{BMM},K=5)}(R) - E^{\mathrm{SQD}}(R)$ and $\Delta E_{K=1} = E^{\mathrm{CSQD}}_{(K=1)}(R) - E^{\mathrm{SQD}}(R)$, while the lower panel reports the RHF determinant weight $|c_{\mathrm{RHF}}|^2$ extracted from the HCI wave function as an indicator of the breakdown of a single-reference description.
Consistent with the decrease in $|c_{\mathrm{RHF}}|^2$ with increasing $R$, both $\Delta E_{K=1}$ and $\Delta E_{K=5}$ trend downward.
In the strongly correlated regime, $\Delta E_{K=1}$ is negative for all but one tested geometry, with the only exception at $R=2.0~\mathrm{\AA}$.
This suggests that part of the CSQD gain already arises in the single-cluster limit.
At the same time, $\Delta E_{K=5}$ remains lower than $\Delta E_{K=1}$ throughout the same region, indicating that cluster-adaptive recovery provides an additional benefit once the wave function exhibits pronounced multireference character.

We now quantify the robustness of this behavior across clustering models and the tested range of $K=2$--$5$ at $\dim(\mathcal{V})=4\times 10^{6}$ ($d_{\max}=2000$).
In the strongly correlated (stretched-bond) regime ($R\ge 1.5R_e$), CSQD yielded lower variational energies ($\Delta E<0$) in all 72 tested BMM cases and in 71 out of 72 tested $k$-modes cases.
The 72 cases correspond to 18 stretched-bond geometries evaluated at four distinct values of $K\in\{2,3,4,5\}$.
For BMM, the energy difference spanned $\Delta E\in[-15.67,-1.21]~\mathrm{mHa}$, with a median and mean of $-8.77$ and $-9.33~\mathrm{mHa}$, respectively.
For $k$-modes, the energy difference spanned $\Delta E\in[-15.95,+1.19]~\mathrm{mHa}$, with a median and mean of $-7.68$ and $-7.45~\mathrm{mHa}$, respectively.
The single instance where SQD yielded a lower energy occurred at $K=2$ and $R=2.8~\mathrm{\AA}$ ($\Delta E=+1.19~\mathrm{mHa}$).
Overall, this highly consistent advantage in the strongly correlated regime (143 out of 144 tested cases) indicates that the benefit of CSQD is robust across clustering models and across the tested range of $K$.

By contrast, in the weakly correlated regime ($R<1.5R_e$), SQD yielded lower variational energies than CSQD in the majority of settings.
At $d_{\max}=2000$ across $K=2$--$5$, SQD yielded lower energies in 33/40 BMM cases and 38/40 $k$-modes cases.
The corresponding energy differences spanned $\Delta E\in[-0.77,+3.45]~\mathrm{mHa}$ for BMM and $\Delta E\in[-0.53,+4.57]~\mathrm{mHa}$ for $k$-modes, with medians and means of $+1.16$ and $+1.17~\mathrm{mHa}$ (BMM) and $+1.72$ and $+1.91~\mathrm{mHa}$ ($k$-modes), respectively.
This behavior is consistent with a near-single-reference regime, where a single global reference occupancy vector already provides an effective guide for particle-number recovery.

\subsection{$\mathrm{[2Fe\!-\!2S]}$ Cluster: Strong Correlation Benchmark}
Iron--sulfur (Fe--S) clusters are stringent benchmarks for studying strong correlation, with dense, highly multireference low-energy manifolds.
We studied the synthetic $\mathrm{[Fe_2S_2(SCH_3)_4]^{2-}}$ complex (abbreviated $\mathrm{[2Fe\!-\!2S]}$)~\cite{fes} using the same localized-orbital active-space Hamiltonian as in Ref.~\cite{sqd}: an effective $(30e,20o)$ model spanning Fe(3d) and S(3p) orbitals derived from a BP86 calculation~\cite{bb861,bb862} with a TZP-DKH basis~\cite{tzp} and a spin-free X2C treatment~\cite{x2c1,x2c2}.

Crucially, we did not perform any new quantum sampling for this benchmark.
Instead, we reused the measurement dataset~\cite{sqd_repository} released with Ref.~\cite{sqd} as input to both SQD and CSQD.
The dataset consists of $\sim 2.46\times 10^{6}$ computational-basis measurement outcomes generated using the SQD sampling protocol.
Details of dataset generation and mitigation are provided in the Methods section.

For both SQD and CSQD, we swept the matched final projected-subspace dimension $\dim(\mathcal{V})=d_{\max}^2$ with $d_{\max}\in\{1000,1500,2000\}$.
For CSQD, we swept the clustering model (BMM or $k$-modes) and the number of clusters $K\in\{2,3,4,5\}$, and additionally performed $K=1$ runs as a single-cluster ablation of the CSQD workflow.
In the $\mathrm{[2Fe\!-\!2S]}$ benchmark, the released sample set was sufficiently large that both SQD and CSQD achieved the target dimension $\dim(\mathcal{V})=d_{\max}^2$ already in the first iteration.
We therefore ran the self-consistent loop for at most 10 iterations for both methods.
Each iteration used $B=10$ independent batches.
We drew sufficient strings in each batch so that, after merging newly sampled and carry-over strings and removing duplicates, the unique single-spin pool in each spin sector was capped at $d_{\max}$ strings and, in practice, reached this cap.

Table~\ref{tab:2fe2s_sweep} summarizes the final variational energies for SQD and CSQD across all tested settings for each matched final projected-subspace dimension $\dim(\mathcal{V})=d_{\max}^2$ after 10 self-consistent iterations.
At matched $d_{\max}\in\{1000,1500,2000\}$, CSQD yielded lower variational energies than SQD for all tested clustering choices.
The corresponding energy reductions ranged from $20.64$ to $24.17~\mathrm{mHa}$ at $\dim(\mathcal{V})=1.00\times10^{6}$, from $14.22$ to $33.58~\mathrm{mHa}$ at $\dim(\mathcal{V})=2.25\times10^{6}$, and from $30.27$ to $45.53~\mathrm{mHa}$ at $\dim(\mathcal{V})=4.00\times10^{6}$.
The $K=1$ single-cluster ablation also improved upon SQD at all three projected-subspace dimensions, with energy reductions of $19.25$, $12.23$, and $18.98~\mathrm{mHa}$, respectively.
These results indicate that part of the gain is already present in the single-cluster limit, while additional improvements are obtained when cluster-adaptive recovery is used.

\begin{table}[t]
\centering
\small
\setlength{\tabcolsep}{2.2pt}
\renewcommand{\arraystretch}{1.05}
\caption{\textbf{$\mathbf{[2Fe\!-\!2S]}$ ground-state energy estimates from SQD and CSQD under identical quantum samples after 10 self-consistent iterations.}
Results are reported for $d_{\max}\in\{1000,1500,2000\}$ (with final $\dim(\mathcal{V})=d_{\max}^2$).
Energies are in $E_h$ and $\Delta E \equiv E^{\mathrm{CSQD}}-E^{\mathrm{SQD}}$ is in mHa.
Negative $\Delta E$ indicates an improvement by CSQD.
Values are rounded to 5 decimal places in E$_h$ and 2 decimal places in mHa.}
\label{tab:2fe2s_sweep}
\begin{tabular}{@{}c@{\hspace{1pt}}l r c r c r c @{}}
\toprule
 &  & \multicolumn{2}{c}{$\dim(\mathcal{V}) = 1.00\times10^{6}$}
    & \multicolumn{2}{c}{$\dim(\mathcal{V}) = 2.25\times10^{6}$}
    & \multicolumn{2}{c}{$\dim(\mathcal{V}) = 4.00\times10^{6}$} \\
\cmidrule(lr){3-4}\cmidrule(lr){5-6}\cmidrule(lr){7-8}
\multicolumn{2}{c}{Method}
& $E$ (E$_h$) & $\Delta E$ (mHa)
& $E$ (E$_h$) & $\Delta E$ (mHa)
& $E$ (E$_h$) & $\Delta E$ (mHa) \\
\midrule
\multicolumn{2}{c}{SQD}
& -116.39387 & 0.00
& -116.41088 & 0.00
& -116.43666 & 0.00 \\
\midrule
\multicolumn{2}{c}{CSQD ($K$=1)}
& -116.41312 & -19.25
& -116.42311 & -12.23
& -116.45564 & -18.98 \\
\midrule
\multirow{8}{*}{\rotatebox[origin=c]{90}{CSQD}}
& BMM ($K$=2)
& -116.41751 & -23.65
& -116.42550 & -14.62
& -116.46693 & -30.27 \\
& BMM ($K$=3)
& -116.41804 & -24.17
& -116.42712 & -16.24
& -116.47723 & -40.57 \\
& BMM ($K$=4)
& -116.41728 & -23.41
& -116.42632 & -15.44
& -116.47366 & -37.00 \\
& BMM ($K$=5)
& -116.41599 & -22.13
& -116.44445 & -33.58
& -116.48219 & -45.53 \\
\cmidrule(lr){2-8}
& $k$-modes ($K$=2)
& -116.41531 & -21.44
& -116.43242 & -21.54
& -116.46926 & -32.60 \\
& $k$-modes ($K$=3)
& -116.41494 & -21.07
& -116.43738 & -26.50
& -116.47500 & -38.34 \\
& $k$-modes ($K$=4)
& -116.41545 & -21.58
& -116.42510 & -14.22
& -116.47277 & -36.11 \\
& $k$-modes ($K$=5)
& -116.41451 & -20.64
& -116.42548 & -14.60
& -116.47376 & -37.10 \\
\bottomrule
\end{tabular}
\end{table}

\begin{figure}[t]
\centering
\includegraphics[width=1.0\textwidth]{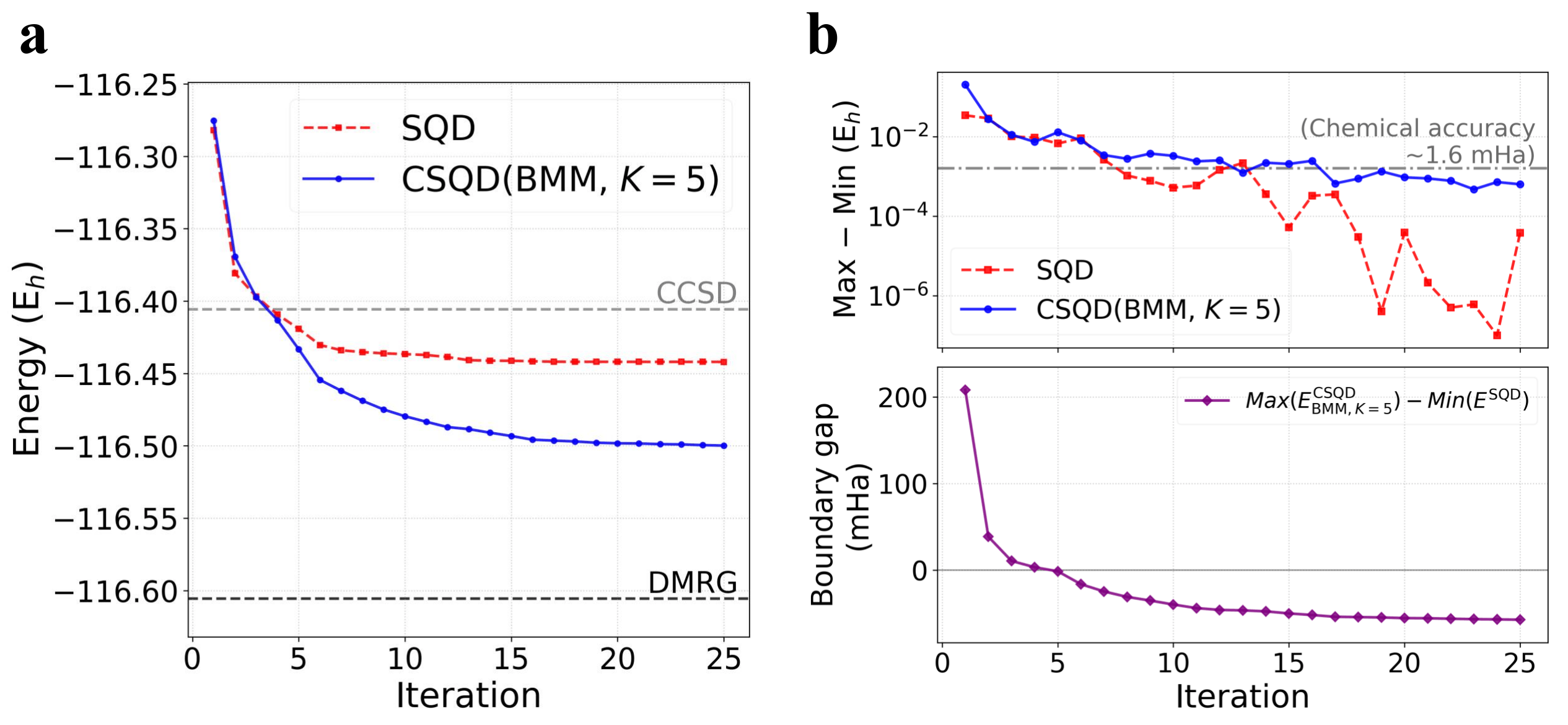}
\caption{\textbf{Extended 25-iteration energy trajectories and batchwise spread diagnostics for $\mathbf{[2Fe\!-\!2S]}$.}
\textbf{(a)} Iteration-wise variational energies for SQD and CSQD (BMM, $K=5$) at $d_{\max}=2000$, corresponding to a final projected-subspace dimension of $\dim(\mathcal{V})=4\times10^{6}$.  
At iteration $t$, the plotted value is the lowest energy among the $B=10$ batches.
Horizontal dashed lines indicate the CCSD and DMRG energies for the same effective active-space Hamiltonian, providing single-reference and high-accuracy classical baselines, respectively. 
\textbf{(b)} Batchwise diagnostics for the same 25-iteration runs. 
The upper panel shows the per-iteration batch-energy range, $\max(E)-\min(E)$, for SQD and CSQD on a logarithmic scale. 
The dash-dotted horizontal line marks chemical accuracy ($1.5936\times10^{-3}\,E_h \approx 1.6$ mHa). 
The lower panel shows the boundary gap, $\max(E^{\mathrm{CSQD}}_{\mathrm{BMM},K=5})-\min(E^{\mathrm{SQD}})$, evaluated across the $B=10$ batches at each iteration and reported in mHa.
Values below zero mean that the CSQD batch-energy interval lies entirely below the SQD batch-energy interval at the same iteration.
}
\label{fig3}
\end{figure}

To assess whether the SQD--CSQD energy gap persists beyond the 10-iteration benchmark sweep, we extended both methods to a common fixed budget of 25 self-consistent iterations.
For this analysis, we selected the largest value of $d_{\max}$ considered, $d_{\max}=2000$, and a representative CSQD setting, BMM with $K=5$.
Fig.~\ref{fig3}a shows that both methods lower the energy rapidly in the initial iterations, but their late-stage behaviors differ markedly. 
By iteration 25, both trajectories were close to a plateau, with SQD reaching $-116.44204~E_h$ and CSQD reaching $-116.49986~E_h$, so that CSQD remained lower by $57.82~\mathrm{mHa}$. 
Notably, the SQD trajectory began to level off after iteration 15, showing only an additional $0.78~\mathrm{mHa}$ decrease between iterations 15 and 25, whereas CSQD continued to decrease by $6.64~\mathrm{mHa}$ over the same interval. 
In Fig.~\ref{fig3}a, the CCSD and density matrix renormalization group (DMRG)~\cite{dmrg} reference energies provide single-reference and high-accuracy classical baselines, respectively. 
The corresponding reference energies are $-116.40566~E_h$ and $-116.60561~E_h$, taken from the data~\cite{sqd_repository} released with Ref.~\cite{sqd}. 

Fig.~\ref{fig3}b provides batchwise diagnostics to test whether the CSQD advantage shown in Fig.~\ref{fig3}a could be driven by isolated low-energy outlier batches. 
The upper panel shows that the per-iteration batch-energy range, $\max(E)-\min(E)$, contracts rapidly for both methods. 
The SQD range first falls below chemical accuracy ($1.5936\times10^{-3}\,E_h \approx 1.6$ mHa) at iteration 8 and remains below it from iteration 14 onward, whereas the CSQD range first falls below chemical accuracy at iteration 13 and remains below it from iteration 17 onward. 
The later contraction of the CSQD range is consistent with the continued decrease in the CSQD variational-energy trajectory over the same interval, whereas the SQD trajectory is already close to a plateau.
Averaged over late iterations 21--25, the mean batch range is $0.0083~\mathrm{mHa}$ for SQD and $0.694~\mathrm{mHa}$ for CSQD, indicating that the residual late-stage batch-to-batch variability is small on a chemically relevant scale for both methods. 
The lower panel shows that the boundary gap, $\max(E^{\mathrm{CSQD}}_{\mathrm{BMM},K=5})-\min(E^{\mathrm{SQD}})$, becomes negative at iteration 5 and then decreases monotonically, reaching $-57.19~\mathrm{mHa}$ at iteration 25.
Thus, from iteration 5 onward, the full CSQD batch-energy interval lies below the corresponding SQD interval at every iteration in this representative run. 
Together, these results indicate that the separation seen in Fig.~\ref{fig3}a reflects a systematic downward shift of the CSQD batch-energy distribution rather than a gain arising from favorable outlier batches.
Exact iteration-resolved batch-energy summary statistics for the 25-iteration runs are reported in the Supplementary Information.

We next analyze the final reference occupancy vectors and string pools after 25 iterations.
Each reference occupancy vector $\mathbf n$ is normalized to satisfy the electron-number constraint in a given spin sector, $\sum_p n_p = N_\sigma$, and represents the cluster-averaged occupations of the orbitals.
In this sense, $\mathbf n$ can be interpreted as a centroid in occupation space that summarizes the configurations assigned to that cluster.
We quantify the separation between two such occupancy vectors using one-half of the Manhattan ($L_1$) distance,
\begin{equation}
\eta_{ij}\equiv \frac{1}{2}\left\lVert \mathbf n^{(i)}-\mathbf n^{(j)}\right\rVert_{1}
= \frac{1}{2}\sum_{p}\left|n^{(i)}_{p}-n^{(j)}_{p}\right|.
\label{eq:eta}
\end{equation}
Because $\mathbf n$ is expressed in electron units, $\eta_{ij}$ is also measured in electrons and captures the total occupation weight that must be redistributed among orbitals to transform one set of orbital occupations into another.
For SQD, we used the $\alpha/\beta$-averaged occupations,
$\mathbf n^{\mathrm{SQD}} \equiv (\mathbf n^{\mathrm{SQD}}_{\alpha}+\mathbf n^{\mathrm{SQD}}_{\beta})/2$,
which is appropriate in the $S_z=0$ sector where $\mathbf n^{\mathrm{SQD}}_{\alpha}=\mathbf n^{\mathrm{SQD}}_{\beta}$ in the ideal spin-inversion-symmetric limit.

\begin{figure}[t]
\centering
\includegraphics[width=1.0\textwidth]{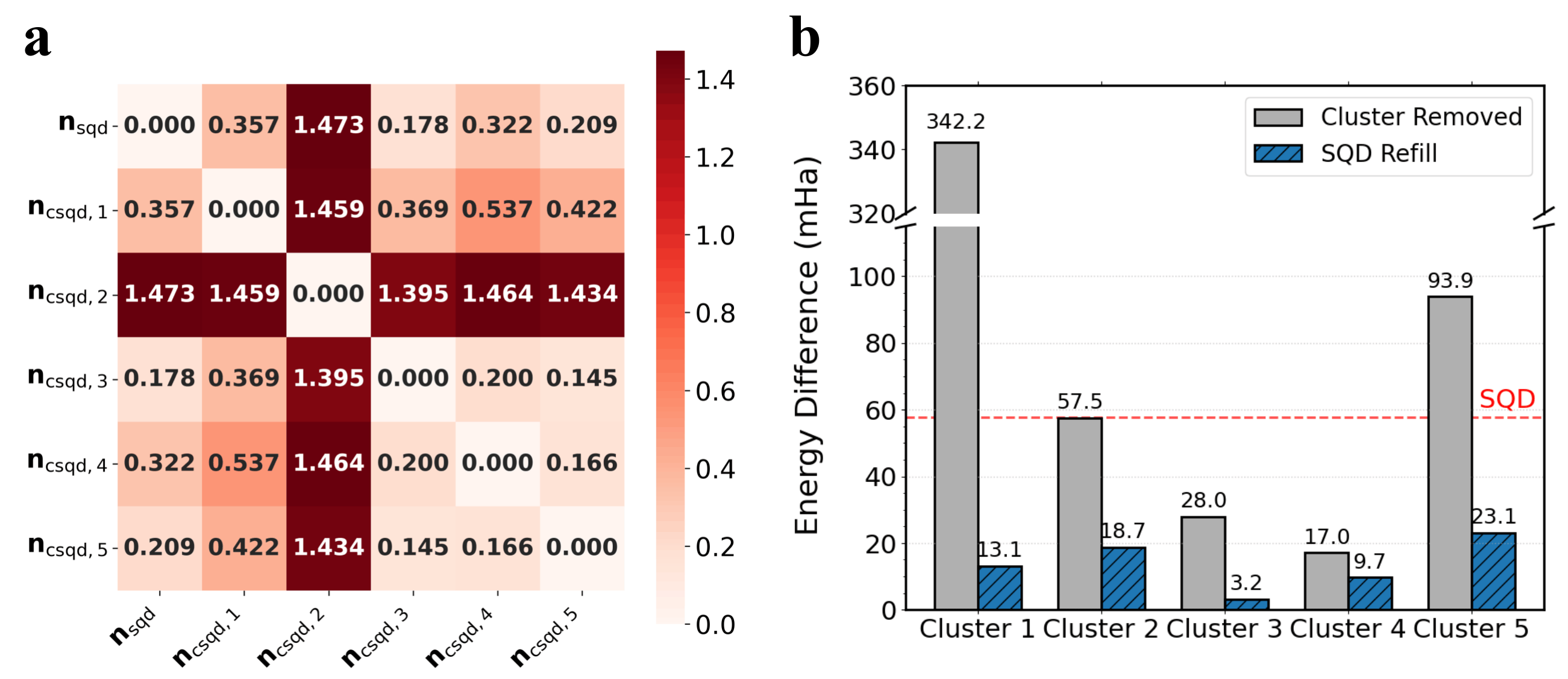}
\caption{
\textbf{Reference-vector analysis and post hoc removal/refill diagnostics for the 25-iteration $\mathbf{[2Fe\!-\!2S]}$ SQD and CSQD runs.}
Results are shown at $d_{\max}=2000$, with CSQD using BMM clustering and $K=5$.
\textbf{(a)} Heatmap of one-half the Manhattan distance ($L_1$) computed for all pairs among the SQD global reference occupancy vector $\mathbf{n}_{\mathrm{SQD}}$ and the CSQD cluster-specific reference occupancy vectors $\mathbf{n}^{(k)}_{\mathrm{CSQD}}$.
\textbf{(b)} Variational energy penalties relative to the full CSQD result. 
Bars denote the energy increase after removing each cluster's assigned strings (Cluster Removed) and the remaining residual penalty after refilling the single-spin budget to 2000 strings using SQD-selected configurations (SQD Refill).
}
\label{fig4}
\end{figure}

Fig.~\ref{fig4}a visualizes the pairwise-distance matrix $\eta_{ij}$ for the reference occupancy vectors, including the SQD global reference and the CSQD cluster-specific references (BMM, $K=5$).
The matrix shows that the final CSQD reference vectors are nonuniform in occupation space rather than collapsing toward a single global pattern.
More specifically, Cluster~2 is the most distinct cluster, with $\eta = 1.473$ relative to the SQD reference and $\eta = 1.395$--$1.464$ relative to the other CSQD clusters.
Cluster~1 lies at an intermediate distance from the SQD reference ($\eta = 0.357$) and from the other CSQD clusters ($\eta = 0.369$--$1.459$).
Clusters~3--5 form a relatively compact subgroup whose members remain closer to the SQD reference ($\eta = 0.178$, $0.322$, and $0.209$, respectively) and to one another ($\eta = 0.145$--$0.200$ among Clusters~3--5).

To assess the role of each cluster in the final 25-iteration CSQD solution, we performed a post hoc cluster-removal-and-refill analysis.
Because the reference vectors evolved during the self-consistent loop and cluster memberships were accumulated across iterations rather than redefined solely by the final references (see Methods), the final 2000 CSQD single-spin strings were not directly labeled by the final clusters. 
We therefore assigned them to the nearest final reference vector according to the $L_1$ distance.
This yielded 847, 700, 79, 119, and 255 assigned strings for Clusters~1--5, respectively.
We first evaluated the effect of each cluster by removing its assigned strings and diagonalizing the Hamiltonian projected onto the resulting reduced space. 
However, simple removal alone cannot determine whether the energy increase arises from the loss of a specific sector or merely from the reduced dimension of the projected space. 
To resolve this ambiguity, we introduced a budget-matched control alongside the raw-removal test. 
After removing a given cluster, we restored the single-spin pool to exactly 2000 strings. 
This deficit was refilled first by reinserting removed strings that also belonged to the final SQD pool (accounting for 548, 278, 56, 25, and 134 strings for Clusters~1--5, respectively), and then by adding the highest-importance SQD strings ranked according to the same metric used in the CSQD carry-over step (see Methods).

Removing the assigned strings without refill raised the variational energy by $342.25, 57.54, 27.99, 17.04,$ and $93.86~\mathrm{mHa}$ for Clusters~1--5, respectively, relative to the full CSQD result.
After the single-spin budget was restored to 2000 through SQD-guided refill, these penalties dropped to $13.14, 18.65, 3.18, 9.68,$ and $23.15~\mathrm{mHa}$.
Although these penalties were smaller than the full SQD--CSQD energy gap of $57.82~\mathrm{mHa}$, they show that the full CSQD result is not completely recovered by replacing the removed content with SQD-selected strings under the same final single-spin budget.
The results of this removal-and-refill analysis are summarized in Fig.~\ref{fig4}b.

We next analyze jointly the cluster-resolved distance metrics in Fig.~\ref{fig4}a and the cluster-resolved residual penalties in Fig.~\ref{fig4}b.
Cluster~2 is the most distant sector from SQD ($\eta = 1.473$) and still exhibits a substantial residual penalty after refill ($18.65~\mathrm{mHa}$), indicating that a single global SQD reference does not adequately represent this sector.
By contrast, Cluster~3 is the sector closest to the SQD reference ($\eta = 0.178$) and exhibits the smallest residual penalty after refill ($3.18~\mathrm{mHa}$).
This is consistent with the SQD global reference already spanning much of the relevant configuration space.
However, Cluster~5 exhibits contrasting behavior.
It remains relatively close to the SQD reference ($\eta = 0.209$) yet exhibits the largest residual penalty ($23.15~\mathrm{mHa}$).
We interpret this discrepancy as suggesting that, in strongly correlated regimes, the determinant support can remain dispersed, meaning that the relevant configurations may not be tightly concentrated even around the centroid.
A single global reference captures only the average position of this broad distribution and may therefore fail to encompass its full extent.
Under such conditions, it can become advantageous to partition this dispersed support into multiple local sectors, each guided by its own reference center.


\section{Discussion}\label{discussion}

In this work, we introduced CSQD to overcome a limitation of SQD in strongly correlated systems. 
In standard SQD, particle-number recovery is guided by a single global reference occupancy vector. 
When the determinant distribution is dispersed or heterogeneous, as is often the case in strongly correlated systems, this reference can become effectively averaged over distinct sectors. 
CSQD addresses this issue through cluster-adaptive recovery built on single-spin strings with cluster-specific reference occupancy vectors that are updated self-consistently.

The benchmark results comparing SQD and CSQD revealed a clear dependence on the correlation regime. 
In the weakly correlated region of the $\mathrm{N_2}$ dissociation curve, where the state retains predominantly single-reference character, SQD remained competitive. 
In the strongly correlated, bond-stretched region of the $\mathrm{N_2}$ dissociation curve, as well as in the more stringent $\mathrm{[2Fe\!-\!2S]}$ benchmark, CSQD systematically yielded lower variational energies than SQD. 
Taken together, these results indicate that CSQD is advantageous in regimes with multireference character.

A practical question, however, is whether these gains come at the cost of substantially increased classical overhead. 
Compared to SQD, CSQD introduces additional classical work, including clustering and importance evaluation at the single-spin-string level.
Even so, CSQD occasionally required wall-clock times comparable to, or even slightly smaller than, those of SQD.
For example, in the $\mathrm{[2Fe\!-\!2S]}$ benchmark at $d_{\max}=1000$, SQD required 5.85 h, whereas $k$-modes with $K=5$ required 5.57 h over the same 10 self-consistent iterations. 
Such timing differences more likely reflect variability in Davidson-solver convergence across the projected subspaces. 
Taken together, the wall-clock timings suggest that, despite the added overhead, projected diagonalization remains the main practical runtime bottleneck at matched $d_{\max}$ for the settings studied here. 
Detailed classical wall-clock times are reported in the Supplementary Information.

A key implementation choice in CSQD is to operate on single-spin strings rather than full bitstrings $x=(\mu_\alpha,\mu_\beta)$. 
In the $S_z=0$ singlet-target setting, the configurations $(\mu_\alpha,\mu_\beta)$ and $(\mu_\beta,\mu_\alpha)$ are physically equivalent under spin inversion. 
However, clustering in the full $2N_{\mathrm{MO}}$ bitstring space can separate such pairs when $\mu_\alpha$ and $\mu_\beta$ are far apart in Hamming distance. 
Pooling and clustering single-spin strings treat the two spin sectors on equal footing, naturally preserving this physical equivalence. 

One concern with operating in the single-spin representation is that explicit $\alpha$--$\beta$ pairing information is not retained.
CSQD mitigates this loss by assigning each single-spin string an importance weight derived from the variational solution on the product space $\mathcal D \times \mathcal D$ and using it in the carry-over step. 
Specifically, the importance weight of a single-spin string is obtained by aggregating the squared CI amplitudes of all determinants in which that string appears (see Methods). 
As a result, an important determinant $(\mu_\alpha,\mu_\beta)$ contributes simultaneously to the importance weights of both $\mu_\alpha$ and $\mu_\beta$, increasing the likelihood that both strings are retained in $\mathcal D$. 
When both components are retained, the corresponding determinant $(\mu_\alpha,\mu_\beta)$ is explicitly included in the variational space $\mathcal D \times \mathcal D$ and can contribute to the projected diagonalization. 
Thus, although the single-spin-string workflow does not enforce determinant-level pairing, $\alpha$--$\beta$ correlations are partially preserved through the string-level carry-over scheme.

Part of the advantage of CSQD in strongly correlated regimes appears to arise from the single-spin-string workflow itself.
To examine this effect, we additionally considered the $K=1$ control, which can be regarded as the single-cluster limit of CSQD. 
Compared with SQD, it provides a qualitative indication of the benefit of the single-spin-string workflow, while the comparison between $K=1$ and $K>1$ indicates the additional benefit of cluster-adaptive recovery.

The behavior of the $K=1$ control depended on the correlation regime. 
In the weakly correlated region of the $\mathrm{N_2}$ dissociation curve, SQD tended to yield slightly lower variational energies than the $K=1$ control. 
By contrast, in the strongly correlated region of the $\mathrm{N_2}$ dissociation curve, the $K=1$ control often yielded lower variational energies than SQD, with further improvements obtained for $K>1$. 
The same pattern was observed for the $\mathrm{[2Fe\!-\!2S]}$ cluster, where the $K=1$ control already yielded lower variational energies than SQD and larger gains were obtained for $K>1$.

This contrast is consistent with the differing structure of weakly and strongly correlated wave functions. 
In weakly correlated regimes, the ground state is often dominated by a small number of determinants, and the key structural information can reside in specific pairings $(\mu_\alpha,\mu_\beta)$ that carry large coefficients. 
Because SQD subsamples, recovers, and carries over full determinants, it can preserve this pairing structure efficiently.
In strongly correlated regimes, by contrast, the wave-function weight is distributed over many configurations rather than concentrated in a few dominant pairs. 
In this setting, the importance of a given $\alpha$- or $\beta$-string may not be reflected in any single large pair coefficient, but can instead arise from accumulated contributions across a broad region of configuration space. 
Because CSQD evaluates single-spin-string importance by aggregating contributions over determinants in the projected space, it can better retain such strings during the iterative subspace construction.

A practical limitation of the present implementation is that no single clustering model or choice of $K$ was uniformly optimal across the strongly correlated benchmarks considered here.
In the $\mathrm{[2Fe\!-\!2S]}$ benchmark, for example, the clustering choice yielding the lowest variational energy differed across the values of $d_{\max}$, indicating that the optimal setting is not straightforward to predict a priori.
These observations motivate future work on more systematic selection of the clustering model and hyperparameters to improve the robustness of CSQD.

Beyond such implementation-level refinements, it is also important to clarify how CSQD fits into the broader landscape of SQD-based developments.
CSQD is naturally compatible with existing SQD-based extensions because it targets a different bottleneck in the workflow. 
Variants such as SKQD~\cite{skqd} and PIGen-SQD~\cite{pigen} aim to improve the sampling or configuration-generation stage itself. 
Other directions use SQD outputs in classical workflows, including excited-state extensions~\cite{extsqd}, phaseless auxiliary-field quantum Monte Carlo (ph-AFQMC)~\cite{afqmc}, and quantum embedding~\cite{embedding}. 
CSQD instead refines the recovery step for a fixed set of noisy samples, which makes it complementary to both types of developments and a modular refinement of the broader SQD workflow. 
Systematic benchmarking of such combined workflows is beyond the scope of the present work and is left to future study.

More broadly, Ref.~\cite{sqd} suggested that SQD may extend beyond quantum chemistry to problems whose target states are well approximated by sparse vectors. 
In such settings, the dominant basis support may be sparse yet distributed across multiple distinct regions of configuration space, in which case a single global reference can become less informative. 
The cluster-adaptive recovery principle underlying CSQD may therefore be useful more generally. 
At the same time, the present implementation relies on chemistry-specific ingredients, including single-spin strings and particle-number recovery. 
Extending this idea beyond electronic structure will therefore require problem-adapted choices of representation, clustering features, and recovery objectives.

\section{Methods}\label{methods}
\subsection{Hamiltonian, Mapping, and Determinant Representation}
Our goal is to estimate the ground-state energy and wave function of the electronic Born--Oppenheimer Hamiltonian written in second quantization over a finite orthonormal molecular-orbital basis, as defined in Eq.~\eqref{eq:hamiltonian}.
We denote by $\hat a^\dagger_{p\sigma}$ and $\hat a_{p\sigma}$ the fermionic creation and annihilation operators acting on spin orbital $(p,\sigma)$, and define the number operator $\hat n_{p\sigma}=\hat a^\dagger_{p\sigma}\hat a_{p\sigma}$.
Unless otherwise stated, we use atomic units for electronic-structure quantities: lengths are measured in bohr and energies in hartree.
Geometric parameters (e.g., bond lengths) are reported in angstroms (\AA) by convention.
For nonrelativistic all-electron calculations, the nuclear repulsion energy $E_{\mathrm{nuc}}$, the one-electron integrals $h_{pr}$, and the two-electron integrals $(pr|qs)$ are defined as
\begin{align}
E_{\mathrm{nuc}} &= \sum_{a<b}^{N_{\mathrm{nuc}}} \frac{Z_a Z_b}{\lVert \mathbf R_a-\mathbf R_b\rVert}, \label{eq:Enuc_def}\\
h_{pr} &= \int d\mathbf r\, \phi_p^*(\mathbf r)\left[-\frac12\nabla^2-\sum_{a=1}^{N_{\mathrm{nuc}}}\frac{Z_a}{\lVert \mathbf r-\mathbf R_a\rVert}\right]\phi_r(\mathbf r), \label{eq:hpr_def}\\
(pr|qs) &= \int d\mathbf r_1 \int d\mathbf r_2\,
\frac{\phi_p^*(\mathbf r_1)\phi_r(\mathbf r_1)\phi_q^*(\mathbf r_2)\phi_s(\mathbf r_2)}
{\lVert \mathbf r_1-\mathbf r_2\rVert}, \label{eq:eri_def}
\end{align}
where $N_{\mathrm{nuc}}$, $\mathbf R_a$, and $Z_a$ denote the number of nuclei, their positions, and atomic numbers, respectively.
In scalar-relativistic (spin-free) and/or active-space calculations, the indices $p,r,q,s$ label active-space orbitals, and the quantities $E_{\mathrm{nuc}}$, $h_{pr}$, and $(pr|qs)$ are modified to account for scalar-relativistic effects and/or the effective potential generated by inactive electrons.

We represent many-electron basis states using Slater determinants and map them to qubit computational-basis states via the Jordan--Wigner transformation~\cite{jw}.
In this mapping, each qubit encodes the occupancy of a spin orbital, and computational-basis measurements return bitstrings $x\in\{0,1\}^{2N_{\mathrm{MO}}}$, where $N_{\mathrm{MO}}$ is the number of spatial orbitals.
Each bitstring $x$ corresponds uniquely to a determinant $|x\rangle$ obtained by applying creation operators in a fixed canonical order of spin orbitals according to the occupation pattern (see Eq.~\eqref{eq:determinant}).
The Hamiltonian in Eq.~\eqref{eq:hamiltonian} conserves the numbers of $\alpha$ and $\beta$ electrons.
The number of spin-$\sigma$ electrons in configuration $x$ is thus $N_\sigma(x)=\sum_p x_{p\sigma}$, and $x$ lies in the correct particle sector if it satisfies
\begin{equation}
\label{eq:right_particle_sector}
\big(N_\alpha(x),\,N_\beta(x)\big)=\big(N_\alpha,\,N_\beta\big),
\end{equation}
i.e., the spin-resolved electron counts computed from the bitstring match the target sector of the molecular problem.
In the benchmarks studied in this work, we estimate the ground-state energy in the $S_z=0$ sector (i.e., $N_\alpha=N_\beta\equiv N_\sigma$).

\subsection{CSQD: Clustering-Based Configuration Recovery}
CSQD takes computational-basis samples and constructs compact determinant subspaces for classical projection and diagonalization. 
Algorithm~\ref{alg:csqd} provides a high-level schematic of the CSQD loop, including clustering of single-spin strings, cluster-adaptive particle-number symmetry recovery (pool refinement), and iterative subspace construction with carry-over. We next describe CSQD in detail, including clustering, subsampling, reference-vector updates, and projection--diagonalization steps.

\begin{algorithm}[!h]
\caption{CSQD (high-level)}
\label{alg:csqd}
\begin{algorithmic}[1]
\Require Samples $\mathcal{X}\subset\{0,1\}^{2N_{\mathrm{MO}}}$ obtained from $N_{\mathrm{meas}}$ measurements with empirical probabilities $\Pr(x)$; target $(N_\alpha,N_\beta)$ with $N_\alpha=N_\beta\equiv N_\sigma$ ($S_z=0$).
Params: $K$ (clusters), $B$ (batches per iteration), $S$ (per-batch new-sample budget), $d_{\max}$ (max single-spin string budget), $T$ (max iterations), $\tau$ (carry-over threshold), $(\varepsilon_E,\varepsilon_n)$ (stopping tolerances).
\Ensure Lowest-energy found subspace solution $(E^\star,C^\star,\mathcal D^\star)$ and $\{\mathbf n^{(k)}\}_{k=1}^K$, where $C^\star$ denotes CI coefficients on $\mathcal D^\star\times\mathcal D^\star$.

\Statex \textbf{Preprocess.} Split $x\in\mathcal X$ into single-spin strings $\mu$; pool unique $\mu$ with weights; cluster into $\{\mathcal C_k\}_{k=1}^K$ with weights $w_k$; set $m_k\gets\lceil w_k S\rceil$.
\State Initialize $\mathbf n^{(k)}_{\mathrm{raw}}$ as the weighted sum of string occupancies in $\mathcal C_k$.
\State Postselect particle-number-correct outcomes $x$ and re-pool their single-spin strings.
\State Assign postselected strings to clusters to form $\{\mathcal C_k^C\}$; set $\mathcal C_k^I=\mathcal C_k\setminus \mathcal C_k^C$.
\State $E^\star\gets+\infty$; $\mathcal D^\star\gets \emptyset$; $C^\star\gets \emptyset$.

\Statex \textbf{Setup (t=1).} Use only the postselected correct pools.
\State $\tilde{\mathcal C}_k\gets \mathcal C_k^C$ for all $k$.
\For{$b=1,\dots,B$}
  \State \parbox[t]{\dimexpr\linewidth-3em}{Sample $\mathcal D_{\mathrm{new}}$ by drawing $m_k$ elements from each $\tilde{\mathcal C}_k$; if a cluster has fewer than $m_k$ postselected strings, fill the deficit by sampling synthetic strings guided by $\mathbf n^{(k)}_{\mathrm{raw}}$ under the particle-number constraint.}
  \State \parbox[t]{\dimexpr\linewidth-3em}{Construct $\mathcal D\gets \mathcal D_{\mathrm{new}}$; deduplicate (and, if needed, truncate to $d_{\max}$).}
  \State \parbox[t]{\dimexpr\linewidth-3em}{Solve the projected problem on $\mathrm{span}(\mathcal D\times \mathcal D)$ to obtain $(E,C)$; if $E<E^\star$, update $(E^\star,C^\star,\mathcal D^\star)$.}
\EndFor
\State $E_{\mathrm{prev}}\gets E^\star$; $\mathbf n_{\mathrm{prev}}^{(k)}\gets \mathbf n^{(k)}_{\mathrm{raw}}$.

\For{$t=2,\dots,T$}
  \State \parbox[t]{\dimexpr\linewidth-3em}{Update $\mathbf n^{(k)}$ from $(C^\star,\mathcal D^\star)$ (membership-weighted, coefficient-weighted occupancies with particle-number normalization; see Eq.~\eqref{eq:csqd:nvec_update}).}
  \State \parbox[t]{\dimexpr\linewidth-3em}{Refine pools: for each $k$, map $\mu\in\mathcal C_k^I$ to corrected strings via biased bit flips guided by $\mathbf n^{(k)}$; set $\tilde{\mathcal C}_k\gets \mathcal C_k^C\cup\tilde{\mathcal C}_k^I$.}
  \For{$b=1,\dots,B$}
    \State \parbox[t]{\dimexpr\linewidth-3em}{Sample $\mathcal D_{\mathrm{new}}$ from $\tilde{\mathcal C}_k$ and compute $I_{\mathrm{new}}(\mu)$ (see Eq.~\eqref{eq:csqd:inew}).}
    \State \parbox[t]{\dimexpr\linewidth-3em}{Select carry-over $\mathcal D_{\mathrm{carry}}$ from $\mathcal D^\star$ by thresholding $C^\star$ at $\tau$ and compute $I_{\mathrm{carry}}(\mu)$ (see Eq.~\eqref{eq:csqd:carry_weight}).}
    \State \parbox[t]{\dimexpr\linewidth-3em}{Merge and deduplicate: $\mathcal D \gets \mathcal D_{\mathrm{new}}\cup \mathcal D_{\mathrm{carry}}$; compute $I(\mu)=I_{\mathrm{new}}(\mu)+I_{\mathrm{carry}}(\mu)$ and retain the top $d_{\max}$ strings by $I(\mu)$.}
    \State \parbox[t]{\dimexpr\linewidth-3em}{Solve the projected problem on $\mathrm{span}(\mathcal D\times \mathcal D)$ to obtain $(E,C)$; if $E<E^\star$, update $(E^\star,C^\star,\mathcal D^\star)$.}
  \EndFor
  \If{$|E^\star-E_{\mathrm{prev}}|<\varepsilon_E$ \textbf{and} $\max_k\|\mathbf n^{(k)}-\mathbf n_{\mathrm{prev}}^{(k)}\|_\infty<\varepsilon_n$}
    \State \textbf{break}
  \EndIf
  \State $E_{\mathrm{prev}}\gets E^\star$; $\mathbf n_{\mathrm{prev}}^{(k)}\gets \mathbf n^{(k)}$.
\EndFor
\State \Return $(E^\star,C^\star,\mathcal D^\star,\{\mathbf n^{(k)}\})$.
\end{algorithmic}
\end{algorithm}

\paragraph{Input samples and clustering}
From quantum measurements, we obtain $N_{\mathrm{meas}}$ computational-basis outcomes, collected in a set $\mathcal{X}\subset\{0,1\}^{2N_{\mathrm{MO}}}$.
For an observed outcome $x\in\mathcal{X}$, $x$ is a full bitstring encoding spin-orbital occupations and $\Pr(x)$ denotes its empirical probability (normalized frequency) over $\mathcal{X}$.
We split each $x$ into two single-spin strings,
\begin{equation}
x = \bigl(\mu_\alpha,\, \mu_\beta\bigr),\qquad
\mu_\sigma \in \{0,1\}^{N_{\mathrm{MO}}}\ \ (\sigma\in\{\alpha,\beta\}),
\label{eq:csqd:split}
\end{equation}
and use the single-spin strings $\mu$ as the fundamental objects in CSQD.
We pool the single-spin strings observed across all measurement outcomes, deduplicate them, and form a pooled set $\mathcal{X}_{\mu}$.
For each $\mu\in\mathcal{X}_{\mu}$, we assign a pooled empirical weight $\pi(\mu)$ defined by
\begin{equation}
\pi(\mu)
=\frac12 \sum_{x\in\mathcal{X}} \Pr(x)\Bigl[\mathbf{1}\!\left(\mu_\alpha(x)=\mu\right)+\mathbf{1}\!\left(\mu_\beta(x)=\mu\right)\Bigr],\qquad
\sum_{\mu\in\mathcal{X}_{\mu}}\pi(\mu)=1,
\label{eq:csqd:pi_def}
\end{equation}
where $(\mu_\alpha(x),\mu_\beta(x))$ is the decomposition of $x$ in Eq.~\eqref{eq:csqd:split}, and $\mathbf{1}(\cdot)$ equals 1 if the argument is true and 0 otherwise.

We partition the pooled set $\mathcal{X}_{\mu}$ into $K$ clusters $\{\mathcal C_k\}_{k=1}^K$, and define the within-cluster normalized weight
$\pi_k(\mu)=\pi(\mu)/w_k$ for $\mu\in\mathcal C_k$, where $w_k$ is given in Eq.~\eqref{eq:csqd:wk_mk_def}.
In this work, we employ $k$-modes~\cite{kmodes,kmodes_library} and BMM~\cite{bmm,stepmix} for clustering.
We perform clustering once per geometry and keep cluster assignments fixed throughout the CSQD iterations.
The cluster sampling weight and the per-batch allocation are
\begin{equation}
w_k \;=\; \sum_{\mu\in\mathcal C_k}\pi(\mu),\qquad
m_k \;=\; \left\lceil w_k\,S \right\rceil,
\label{eq:csqd:wk_mk_def}
\end{equation}
where $S$ is the per-batch new-sample budget.
We also define the initial (unnormalized) reference occupancy vector for each cluster as
\begin{equation}
\mathbf n^{(k)}_{\mathrm{raw}} \;=\; \sum_{\mu\in\mathcal C_k} \pi(\mu)\, \mu,
\label{eq:csqd:raw_nvec}
\end{equation}
i.e., the probability-weighted sum of string occupations.

\paragraph{Particle-number postselection}
CSQD proceeds in iterations indexed by $t=1,\dots,T$. At $t=1$, we restrict to measurement samples whose full bitstring $x$ satisfies the target particle numbers,
\begin{equation}
\mathrm{hw}(\mu_\alpha)=\mathrm{hw}(\mu_\beta)=N_\sigma,
\label{eq:csqd:postselect}
\end{equation}
where $\mathrm{hw}(\cdot)$ denotes the Hamming weight. We re-pool the single-spin strings from these postselected samples, map them to the precomputed clusters $\{\mathcal C_k\}$, and form cluster-wise pools of particle-number-correct strings.
For later pool refinement, we also partition each cluster into a correct particle-number subset $\mathcal C_k^C$ and an incorrect subset $\mathcal C_k^I=\mathcal C_k\setminus \mathcal C_k^C$.

\paragraph{Iteration $t=1$: initial subspace construction}
For each cluster $k$, we draw up to $m_k$ particle-number-correct strings without replacement from the postselected pool assigned to $\mathcal C_k$, with probabilities proportional to $\pi_k(\mu)$.
If the number of available postselected strings in cluster $k$ is smaller than $m_k$, we fill the deficit by generating additional particle-number-correct strings guided by $\mathbf n^{(k)}_{\mathrm{raw}}$:
we sample $N_\sigma$ occupied orbital indices without replacement using weights proportional to $\mathbf n^{(k)}_{\mathrm{raw}}$.
We then pool the selected strings across clusters, remove duplicates, and form the initial single-spin string set $\mathcal D$.

\paragraph{Projected diagonalization}
Given a selected set of single-spin strings $\mathcal D$, we form the product-basis $\{|\mu\rangle_\alpha|\nu\rangle_\beta\}_{\mu,\nu\in\mathcal D}$ and define the corresponding projector
\begin{equation}
P_{\mathcal D} \;=\; \sum_{\mu\in\mathcal D}\sum_{\nu\in\mathcal D}
|\mu\rangle_\alpha|\nu\rangle_\beta \, {}_\alpha\langle \mu|\, {}_\beta\langle \nu| .
\label{eq:csqd:proj}
\end{equation}
We then construct the projected Hamiltonian $H_{\mathcal D}=P_{\mathcal D}HP_{\mathcal D}$ and solve for its lowest eigenpair to obtain the projected energy and CI coefficients on $\mathcal D\times\mathcal D$. In practice, we solve the projected eigenproblem using PySCF~\cite{pyscf1, pyscf2} with a Davidson-type iterative eigensolver. 
Following SQD~\cite{sqd}, to target the singlet, we encourage $S^2=0$ using a spin-penalty, i.e.,
\begin{equation}
H_{\mathcal D}\ \mapsto\ H_{\mathcal D} + \lambda\bigl(S^2 - s(s+1)\bigr)^2,\qquad s=0,
\label{eq:csqd:spin_penalty}
\end{equation}
with $\lambda>0$ chosen to suppress spin contamination.
In all CSQD and SQD experiments conducted in this study, we set the spin-penalty parameter $\lambda$ to 0.1.
The reported energies are expectation values of the unpenalized projected Hamiltonian $H_{\mathcal D}$, evaluated with the eigenvector obtained from the spin-penalized diagonalization.
For each iteration, we solve $B$ independent batches and update the current best solution $(E^\star,C^\star,\mathcal D^\star)$ when a batch yields a projected energy lower than $E^\star$.

\paragraph{Cluster membership accumulation}
For each string $\mu$, CSQD maintains a Boolean membership vector $\mathbf M_\mu\in \{0, 1\}^K$ indicating which clusters have produced $\mu$. When constructing each batch, the component corresponding to the sampled cluster is set to $1$; if the same string is selected from multiple clusters, memberships are accumulated. Carry-over strings inherit their previous memberships, and overlaps between carry-over and newly sampled strings are also merged via OR. Thus, once a string is ever associated with a cluster, its membership for that cluster remains $1$, which partially softens hard cluster boundaries.

\paragraph{Self-consistent update of reference vectors}
For iterations $t\ge 2$, the cluster-wise reference occupancy vectors $\mathbf n^{(k)}$ are updated from the current best CI coefficients $C^\star$ on $\mathcal D^\star\times\mathcal D^\star$.
We define a string weight
\begin{equation}
w_\mu \;=\; \sum_{\nu\in\mathcal D^\star}\bigl|C^\star_{\mu\nu}\bigr|^2 \;+\; \sum_{\nu\in\mathcal D^\star}\bigl|C^\star_{\nu\mu}\bigr|^2,
\label{eq:csqd:string_weight}
\end{equation}
and update $\mathbf n^{(k)}$ by a membership-weighted average of occupancies, followed by particle-number normalization:
\begin{equation}
\mathbf n^{(k)} \;=\; N_\sigma\,
\frac{\sum_{\mu\in\mathcal D^\star} M_{\mu,k}\,w_\mu\, \mu}{\left\lVert \sum_{\mu\in\mathcal D^\star} M_{\mu,k}\,w_\mu\, \mu \right\rVert_1},
\label{eq:csqd:nvec_update}
\end{equation}
where $\|\cdot\|_1$ denotes the sum of vector components and $M_{\mu,k}$ denotes the $k$th component of the membership vector $\mathbf M_\mu$.
The normalization enforces $\sum_p n^{(k)}_p=N_\sigma$, so $\mathbf n^{(k)}$ can be interpreted as a cluster-resolved reference occupancy profile.

Using the updated $\mathbf n^{(k)}$, we refine each cluster pool by correcting particle-number-incorrect strings in $\mathcal C_k^I$ to produce particle-number-correct candidates, yielding a refined pool
$\tilde{\mathcal C}_k=\mathcal C_k^C\cup\tilde{\mathcal C}_k^I$,
where $\tilde{\mathcal C}_k^I$ denotes the set of corrected strings obtained from $\mathcal C_k^I$ by the procedure below.
For a string $\mu \in \mathcal C_k^I$, we define a flip score for orbital $p$ as
\begin{equation}
d_p \;=\; |\mu_p - n^{(k)}_p|.
\label{eq:csqd:flip_score}
\end{equation}
We use the same modified ReLU weighting as in Ref.~\cite{sqd}:
\begin{equation}
ReLU(d) \;=\;
\begin{cases}
\displaystyle \delta_0\,\frac{d}{f}, & d\le f,\\[6pt]
\displaystyle \delta_0 + (1-\delta_0)\,\frac{d-f}{1-f}, & d>f,
\end{cases}
\qquad
f=\frac{N_\sigma}{N_{\mathrm{MO}}}.
\label{eq:csqd:relu}
\end{equation}
In all experiments, we set $\delta_0=0.01$.
Let $\Delta_{hw}=\mathrm{hw}(\mu)-N_\sigma$. If $\Delta_{hw}>0$, we select $\Delta_{hw}$ occupied positions to flip $1\to 0$; if $\Delta_{hw}<0$, we select $|\Delta_{hw}|$ unoccupied positions to flip $0\to 1$. In both cases, positions are sampled without replacement with probabilities proportional to $ReLU(d_p)$, and the resulting corrected strings are pooled into $\tilde{\mathcal C}_k^I$.

\paragraph{Subsampling, carry-over, and subspace truncation for $t\ge 2$}
At each batch, we draw $m_k$ strings from each refined pool $\tilde{\mathcal C}_k$ without replacement to form a new candidate set $\mathcal D_{\mathrm{new}}$.
We assign each $\mu\in\mathcal D_{\mathrm{new}}$ a subsampling-derived weight $I_{\mathrm{new}}(\mu)$ as follows: cluster-normalize sampling probabilities, scale by the cluster allocation $m_k/\sum_{\ell=1}^K m_\ell$, and sum weights upon deduplication. Equivalently,
\begin{equation}
I_{\mathrm{new}}(\mu)\;=\;\sum_{k:\,\mu\in\mathcal D_{\mathrm{new}}^{(k)}} \hat\pi_k(\mu)\,\frac{m_k}{\sum_{\ell=1}^K m_\ell},
\label{eq:csqd:inew}
\end{equation}
where $\hat\pi_k(\mu)$ denotes the sampling probability used within the refined pool $\tilde{\mathcal C}_k$, normalized so that $\sum_{\mu\in\tilde{\mathcal C}_k}\hat\pi_k(\mu)=1$, after aggregating and renormalizing weights following correction and deduplication.

To stabilize the subspace across iterations, we also carry over strings from the current best subspace $\mathcal D^\star$ using a threshold $\tau$ on the CI coefficients:
\begin{equation}
\mathcal D_{\mathrm{carry}}
=
\Bigl\{ \mu\in \mathcal D^\star:\ 
\max_{\nu\in\mathcal D^\star}|C^\star_{\mu\nu}|>\tau\ \ \text{or}\ \ \max_{\nu\in\mathcal D^\star}|C^\star_{\nu\mu}|>\tau
\Bigr\}.
\label{eq:csqd:carry_set}
\end{equation}
Each carry-over string is assigned the carry-over importance
\begin{equation}
I_{\mathrm{carry}}(\mu)\;=\;\sum_{\nu\in\mathcal D^\star}|C^\star_{\mu\nu}|^2,
\qquad \mu\in\mathcal D_{\mathrm{carry}}.
\label{eq:csqd:carry_weight}
\end{equation}
In the $S_z=0$ sector of a spin-free Hamiltonian with identical $\alpha/\beta$ string pools, the $\alpha$- and $\beta$-marginal weights are equivalent up to an overall factor in the ideal spin-inversion-symmetric limit. 
Hence, we use the $\alpha$-marginal $I_{\mathrm{carry}}(\mu)=\sum_{\nu\in\mathcal D^\star}\left|C^{\star}_{\mu\nu}\right|^{2}$ without loss of generality.
We then merge $\mathcal D_{\mathrm{new}}$ and $\mathcal D_{\mathrm{carry}}$, deduplicate identical strings, and define the final importance
\begin{equation}
I(\mu)\;=\; I_{\mathrm{carry}}(\mu)\;+\;I_{\mathrm{new}}(\mu).
\label{eq:csqd:imp_merge}
\end{equation}
Finally, we retain the top $d_{\max}$ strings by $I(\mu)$ to form the working set $\mathcal D$ for the projected diagonalization step.
As in the $t=1$ case, this rank-based truncation preferentially retains carry-over strings without imposing a hard constraint. Since the physical weights $I_{\mathrm{carry}}(\mu)$ derived from the wave function coefficients are typically orders of magnitude larger than the statistical sampling weights $I_{\mathrm{new}}(\mu)$, this scheme promotes the prioritization of chemically significant strings during truncation.

\paragraph{Stopping criterion}
We terminate the iterations when (i) the improvement in the best energy is below $\varepsilon_E$ and (ii) the maximum change in the reference vectors satisfies
\begin{equation}
\max_{k,p} |n^{(k)}_{p}-n^{(k)}_{\mathrm{prev},p}| < \varepsilon_n.
\label{eq:csqd:stop}
\end{equation}
The hyperparameters $(K,B,S,d_{\max},T,\tau,\varepsilon_E,\varepsilon_n)$ and system-specific settings are reported in the corresponding experimental subsections.

\subsection{$\mathrm{N_2}$ Dissociation—Computational Details}
\paragraph{Molecular Hamiltonian and active-space definition.}
We studied the dissociation curve of $\mathrm{N_2}$ by scanning the internuclear separation $R$ from $0.7$ to $3.4~\mathrm{\AA}$ in steps of $0.1~\mathrm{\AA}$ (28 geometries).
All electronic-structure quantities were computed in the cc-pVDZ basis using RHF as the reference.
We froze the two lowest-energy core spatial orbitals, yielding an active space of 26 spatial orbitals and 10 active electrons with $(N_\alpha,N_\beta)=(5,5)$.
For each geometry, we constructed the active-space one- and two-electron integrals from the RHF reference using the frozen-core CASCI Hamiltonian construction.
We also computed the associated constant energy shift (nuclear repulsion plus frozen-core contribution) and used it throughout the SQD/CSQD post-processing.
Unless otherwise stated, all $\mathrm{N_2}$ energies reported in this section include this constant shift.

\begin{figure}[t]
\centering
\includegraphics[width=1.0\textwidth]{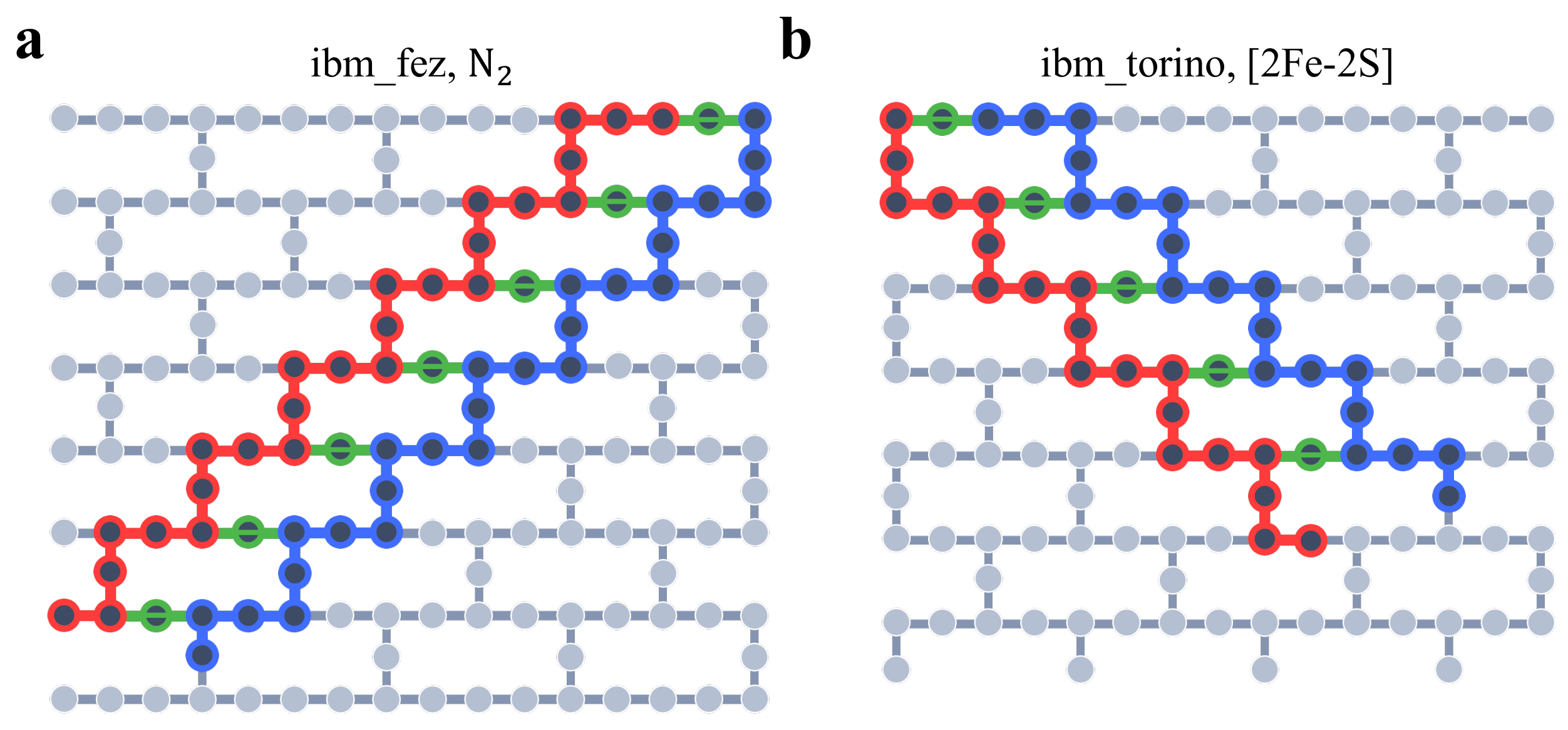}
\caption{\textbf{Hardware mapping on IBM quantum processors.}
Device schematics showing the qubit layouts and role assignments used in this work:
\textbf{(a)} $\mathrm{N_2}$ experiment on \texttt{ibm\_fez} and
\textbf{(b)} $\mathrm{[2Fe\!-\!2S]}$ cluster experiment on \texttt{ibm\_torino}.
Qubits are colored by role: $\alpha$ spin-orbitals (red), $\beta$ spin-orbitals (blue), and auxiliary qubits (green).}
\label{fig5}
\end{figure}

\paragraph{Quantum circuit construction and sampling.}
We followed the circuit-design protocol of Ref.~\cite{sqd} to generate measurement samples for $\mathrm{N_2}$.
We used the LUCJ ansatz~\cite{lucj}, a derivative of the UCJ ansatz~\cite{ucj}. 
The UCJ ansatz has the general form
\begin{equation}
\lvert\Psi\rangle
=\prod_{l=1}^{L} e^{K_{l}} e^{iJ_{l}} e^{-K_{l}} \lvert x_{\mathrm{RHF}} \rangle.
\label{eq:ucj_general}
\end{equation}
In this expression, $K_{l} = \sum_{p,r,\sigma} K^{(l)}_{pr}\,\hat{a}^{\dagger}_{p\sigma}\hat{a}_{r\sigma}$ is a one-body generator, and
$J_{l} = \sum_{p,r,\sigma,\tau} J^{(l)}_{p\sigma,r\tau}\,\hat{n}_{p\sigma}\hat{n}_{r\tau}$ is a number-diagonal Jastrow operator.
We set $L=1$ to limit circuit depth while retaining flexibility via an additional final orbital rotation appended at the end.
The sampled state was thus
\begin{equation}
\lvert\Psi\rangle
= e^{-K_{2}} e^{K_{1}} e^{iJ_{1}} e^{-K_{1}} \lvert x_{\mathrm{RHF}}\rangle,
\label{eq:lucj_truncated}
\end{equation}
where $K_{1}$ and $J_{1}$ correspond to the single ($L=1$) UCJ layer in Eq.~\eqref{eq:ucj_general}, and $K_{2}$ is an additional one-body rotation applied after the correlated layer.
The circuit parameters were obtained from a restricted closed-shell CCSD calculation without additional parameter optimization.

To impose the connectivity-based locality restriction used in the LUCJ construction, we restricted the Jastrow interaction pairs as follows:
(i) for same-spin terms, we retained only nearest-neighbor pairs $(p,p{+}1)$ for $p=0,\dots,24$, and
(ii) for opposite-spin terms, we retained on-site pairs $(p,p)$ for $p \in \{0,4,8,12,16,20,24\}$,
for which the corresponding $\alpha$ and $\beta$ qubits are separated by a single intermediate qubit
(graph distance two) on the heavy-hex topology of IBM devices in Fig.~\ref{fig5}a.
We constructed the LUCJ circuit using the \texttt{ffsim} library~\cite{ffsim}.

Quantum sampling was performed on the IBM Heron r2 backend \texttt{ibm\_fez}.
The 26-orbital active space was encoded using 52 qubits (two spin sectors).
To accommodate a fixed qubit layout on the heavy-hex topology while reducing routing overhead, the sampling circuits occupied 59 device qubits in total, consisting of the 52 spin-orbital qubits and 7 additional routing/auxiliary qubits.
The layout and the assignment of $\alpha$ and $\beta$ spin-orbitals to device qubits are shown in Fig.~\ref{fig5}a.
With this fixed layout, transpilation to the backend native gate set yielded a sampling circuit containing 1876 two-qubit gates.

\paragraph{Error mitigation and reset-mitigation postselection.}
Each geometry was sampled with a total of $3\times 10^{5}$ shots.
We applied dynamical decoupling~\cite{dd} using the XpXm sequence and measurement twirling~\cite{twirling} with 32 randomizations and 9375 shots per randomization.
To mitigate errors arising from imperfect qubit reset, we employed the reset-mitigation scheme demonstrated in Ref.~\cite{sqd}.
This protocol inserts a measurement instruction on all qubits immediately prior to state preparation and discards shots for which the initially measured bitstring differs from the state $0\cdots 0$.
The corresponding final measurement outcomes were then used to form the empirical sample distribution for SQD and CSQD.
After reset-mitigation filtering, we retained on average $\sim 1.08\times 10^{5}$ measurement outcomes per geometry.

\paragraph{SQD and CSQD settings.}
For the SQD baseline, we followed Ref.~\cite{sqd} and utilized the \texttt{qiskit-addon-sqd} package~\cite{qiskit-addon-sqd}.
We set the number of batches per iteration to $B=10$.
From the second recovery iteration onward, $S_{\mathrm{full}}=d_{\max}$ full spin-orbital bitstrings (i.e., 52-bit $\alpha\beta$ measurement outcomes) were drawn for each batch from the corrected full-string pool.
These samples were augmented with carry-over configurations at the level of full spin-orbital bitstrings.
The resulting pool was then split into $\alpha$ and $\beta$ strings and truncated to at most $d_{\max}$ single-spin strings per spin sector prior to the projected diagonalization.
The maximum number of configuration-recovery iterations was set to $T_{\mathrm{SQD}}=11$. We imposed spin-inversion symmetry by symmetrizing the single-spin-string pools (i.e., $\mathcal D_\alpha=\mathcal D_\beta$).
SQD constructed each batch at $t=1$ using only the postselected particle-number-correct full-string pool. 
This often kept the initial projected-subspace dimension per batch well below $d_{\max}^2$. 
We therefore allowed one additional recovery iteration for SQD relative to CSQD.

For CSQD, we used the same target particle number and batching ($B=10$), with a per-batch new-sample budget of $S_{\mathrm{spin}}=d_{\max}$ single-spin strings and a maximum of $T_{\mathrm{CSQD}}=10$ iterations.
To keep the effective variational dimension budget comparable between SQD and CSQD throughout the recovery iterations, we set $S_{\mathrm{full}}=S_{\mathrm{spin}}=d_{\max}$ so that, after splitting and truncation, each spin-sector pool typically remained close to or equal to $d_{\max}$.
We swept the number of clusters $K\in\{2,3,4,5\}$ and considered two clustering methods: weighted $k$-modes clustering (Huang initialization)~\cite{kmodes,kmodes_library} and BMM~\cite{bmm}.
For both methods, the best fit was selected from 100 random initializations.
Specifically, for BMM we retained the run with the largest final lower bound on the weighted log-likelihood, whereas for weighted $k$-modes we retained the run with the smallest weighted clustering cost (sum of distances to the cluster modes).
For the representative setting $d_{\max}=2000$, we additionally performed $K=1$ single-cluster control runs.
In this setting, all pooled single-spin strings were assigned to a single cluster, while the rest of the CSQD workflow and hyperparameters remained unchanged.

For both SQD and CSQD, we truncated the diagonalization subspace by limiting each spin sector to at most $d_{\max}$ single-spin strings.
We considered $d_{\max}\in\{1000,1500,2000\}$, corresponding to a maximum projected dimension of $d_{\max}^{2}$.
We reported results for $d_{\max}=2000$ in the main text, while other values are provided in the Supplementary Information.
Carry-over between iterations used a coefficient threshold of $\tau=10^{-4}$.
Convergence was assessed using $(\varepsilon_{E},\varepsilon_{n})=(10^{-8},10^{-5})$,
where $\varepsilon_{E}$ denotes the threshold for energy improvement and $\varepsilon_{n}$ denotes the maximum change in orbital occupancies (infinity norm).
Configuration recovery utilized the modified-ReLU flip weighting with $\delta_{0}=0.01$ (see Eq.~\eqref{eq:csqd:relu}).
Projected diagonalizations targeted the singlet sector (see Eq.~\eqref{eq:csqd:spin_penalty}).

All classical post-processing tasks were executed as single-node jobs on compute nodes with identical hardware specifications (56-core Intel Xeon Gold 6348R with 1~TB RAM).

\subsection{$\mathrm{[2Fe\!-\!2S]}$ Cluster—Computational Details}
\paragraph{Molecular Hamiltonian and active-space definition.}
We considered the synthetic $\mathrm{[Fe_2S_2(SCH_3)_4]^{2-}}$ complex (abbreviated $\mathrm{[2Fe\!-\!2S]}$)~\cite{fes}.
Following Ref.~\cite{sqd}, we employed the same localized-orbital active-space Hamiltonian with an effective $(30e,20o)$ active space.
The active orbitals span the Fe(3d) and S(3p) manifold and were obtained from a localized BP86 density functional theory calculation in the TZP-DKH basis with a spin-free X2C Hamiltonian to include scalar relativistic effects.
We worked in the $S_z=0$ sector with $(N_\alpha,N_\beta)=(15,15)$ and targeted the singlet sector ($S^2=0$) in the projected diagonalizations.
The one- and two-electron integrals were taken from the data release~\cite{sqd_repository} accompanying Ref.~\cite{sqd}. We reported energies using the same Hamiltonian convention as Ref.~\cite{sqd}.

\paragraph{Quantum circuit and released measurement dataset.}
We did not perform any new quantum sampling for the $\mathrm{[2Fe\!-\!2S]}$ benchmark.
Instead, we reused the fixed computational-basis measurement dataset~\cite{sqd_repository} released with Ref.~\cite{sqd}, and fed the identical empirical sample distribution to both SQD and CSQD.
The released samples were obtained by executing a truncated LUCJ circuit~\cite{lucj} on an IBM Heron r1 backend (\texttt{ibm\_torino}).
The circuit parameters were obtained from a restricted closed-shell CCSD calculation without additional parameter optimization.
The 20-orbital active space was encoded under the Jordan--Wigner transformation on 40 qubits, while the hardware implementation used 45 qubits, 40 Jordan--Wigner qubits plus 5 auxiliary routing qubits on the heavy-hex connectivity (see Fig.~\ref{fig5}b).
This $\mathrm{[2Fe\!-\!2S]}$ circuit contains 1100 two-qubit gates.
The released dataset contains $2.4576\times 10^{6}$ accepted measurement outcomes for this circuit.

\paragraph{Mitigation and preprocessing.}
The released dataset~\cite{sqd_repository} incorporated the experimental mitigation stack of Ref.~\cite{sqd}, including twirled readout error mitigation (ROEM)~\cite{roem} and dynamical decoupling (DD)~\cite{dd}, as well as reset-mitigation.
The postselected outcomes were used to form the empirical sample distribution that served as the common input to SQD and CSQD.

\paragraph{SQD and CSQD settings.}
Unless otherwise stated, we used the same SQD/CSQD hyperparameters as in the $\mathrm{N_2}$ benchmark.
For both SQD and CSQD, we used $B=10$ independent batches per iteration and swept $d_{\max}\in\{1000,1500,2000\}$.
To keep the effective variational dimension budget comparable throughout the recovery iterations, we set $S_{\mathrm{full}}=S_{\mathrm{spin}}=d_{\max}$ and capped each spin-sector pool at $d_{\max}$ unique single-spin strings.
CSQD swept the clustering model (weighted $k$-modes~\cite{kmodes,kmodes_library} and BMM~\cite{bmm}) and the number of clusters $K\in\{2,3,4,5\}$.
We also performed $K = 1$ single-cluster control runs for all tested $d_{\max}$ values.
Because the released $\mathrm{[2Fe\!-\!2S]}$ sample set was sufficiently large that the postselected pools already saturated $d_{\max}$ at $t=1$, we ran both SQD and CSQD for $T=10$ self-consistent iterations.

For the representative extended $\mathrm{[2Fe\!-\!2S]}$ follow-up, we selected $d_{\max} = 2000$ and BMM with $K = 5$, and continued both SQD and CSQD for $T = 25$ self-consistent iterations under the same batching and truncation settings as in the main benchmark sweep. 

All classical post-processing for this benchmark was performed as single-node jobs using 64 CPU cores on an AMD EPYC 9755 system with 1~TB RAM.

\paragraph{Post hoc removal-and-refill analysis.}
For the post hoc cluster-removal-and-refill analysis, each final CSQD single-spin string was assigned to the nearest final reference vector $\mathbf{n}^{(k)}_{\mathrm{CSQD}}$ based on the $L_1$ distance.
For each cluster $k$, a raw removal energy was first calculated by projecting and diagonalizing the Hamiltonian in a reduced space where all strings assigned to that cluster were excluded. 
To isolate the sector-specific energetic contribution from the effect of simple basis-set truncation, the single-spin pool was restored to exactly $d_{\max} = 2000$ strings through a prioritized refill. 
This was performed first by reinserting removed strings that were also present in the final SQD pool. 
The remaining deficit was then filled by adding the highest-ranked SQD strings, where the importance of each string was evaluated based on the final SQD solution using the metric defined in Eq.~\eqref{eq:csqd:carry_weight}.
The Hamiltonian was then projected onto this refilled space and diagonalized to obtain the refill energy, from which the residual penalty was determined relative to the full CSQD result.

\subsection{Software and computational environment}
LUCJ ansatz circuits were generated using \texttt{ffsim} 0.0.64~\cite{ffsim}.
Circuit compilation and execution were performed using Qiskit 2.2.3 and qiskit-ibm-runtime 0.44.0~\cite{qiskit}.
SQD post-processing used qiskit-addon-sqd 0.12.0~\cite{qiskit-addon-sqd}.
Classical electronic-structure calculations and projected diagonalizations used PySCF 2.11.0~\cite{pyscf1,pyscf2}.
For clustering in CSQD, BMM used \texttt{stepmix} 2.2.3~\cite{stepmix}, and $k$-modes used the \texttt{kmodes} library 0.12.2~\cite{kmodes_library}.

\vspace{0.5cm}

\backmatter

\bmhead{Data availability}
All data supporting the findings of this study are available within the paper, its Supplementary Information, and the Zenodo repository at \url{https://zenodo.org/records/19560451}.

\vspace{0.1cm}

\bmhead{Code availability}
The CSQD source code is available from the AIQF Chem Lab Algorithms page (\url{https://stalwart-cassata-0d784a.netlify.app/tags/algorithm}).
An archived snapshot of the version used to generate the results reported in this study is also available on Zenodo at \url{https://zenodo.org/records/19560451}.

\vspace{0.1cm}

\bmhead{Acknowledgements}
This work was supported by the National Research Foundation of Korea (NRF) grant funded by the Korean government (MSIT) (no. RS-2025-005194288) and also Institute of Information \& Communications Technology Planning \& Evaluation (IITP) grant funded by the Korean government (MSIT) (no. RS-2025-25464788). 
The authors also acknowledge the Urban Big data and AI Institute of the University of Seoul supercomputing resources (\url{http://ubai.uos.ac.kr}) made available for conducting the research reported in this paper.

\vspace{0.1cm}

\bmhead{Author contributions}
B.P. and K.J. conceived the study.
B.P. developed the CSQD methodology.
B.P., S.K., J.S., and J.B. designed the numerical experiments and implemented the software and performed computations.
B.P., S.K., J.S., J.B., D.A., and K.J. analyzed and interpreted the results.
K.J. acquired funding and supervised the project.
B.P. wrote the initial manuscript draft, and all authors reviewed and edited the manuscript.

\vspace{0.1cm}

\bmhead{Competing interests}
The authors declare no competing interests.

\bibliography{sn-bibliography}

\end{document}